\documentclass[letterpaper,twocolumn,10pt]{article}

\usepackage{hyperref}
\usepackage{breakurl}

\usepackage{usenix}

\usepackage[numbers,sort,compress]{natbib}

\usepackage{tightenum}
\usepackage{wrapfig}
\usepackage{pifont}

\usepackage{enumitem}
\usepackage{listings}
\usepackage{latexsym}
\usepackage{adjustbox}
\usepackage{graphicx}
\usepackage{floatflt}
\usepackage{setspace}
\usepackage{algorithm}
\usepackage{amsmath}
\usepackage{xspace}
\usepackage{makecell}
\usepackage[font=small]{caption}

\usepackage[normalem]{ulem}

\renewcommand{\em}{\it}

\newcommand{\comment}[1]{}
\newcommand{\ignore}[1]{}

\newcommand{\boldunderpara}[1]{\noindent{\underline{\textbf{#1}}}}

\newcommand{\myitemit}[1]{\item \textit{#1  }}

\newcommand{\figWidthSix}{1.71in}

\def\cffigure[#1,#2,#3]{
\begin{floatingfigure}{3.5in}
\vspace*{-2mm}
\begin{center}

\includegraphics[width=3in]{#1} 
 
\vspace*{-3mm}\caption[]{#2\vspace*{3ex}}
\label{#3} 
\vspace*{-5mm}
\end{center}
\vspace*{-2mm}
\end{floatingfigure}}

\def\cfigure[#1,#2,#3]{
\begin{figure}
\vspace*{0mm}
\begin{center}

\includegraphics[width=3in]{#1} 
 
\vspace*{-3mm}\caption[]{#2
} \label{#3}
 
\vspace*{-5mm}
\end{center}
\vspace*{-2mm}
\end{figure}}

\def\wfigure[#1,#2,#3]{
\begin{figure*}
\vspace*{0mm}
\begin{center}
 
\includegraphics[width=6in]{#1} 
 
\vspace*{-3mm}\caption[]{#2
} \label{#3}
 
\vspace*{-5mm}
\end{center}
\vspace*{-2mm}
\end{figure*}}

\def\dcfigure[#1,#2,#3,#4,#5,#6]{
{
\begin{figure*}
\vspace*{0.0in}\
\begin{center}
\begin{minipage}[c]{3in}{
\includegraphics[width=3in]{#1} 
\vspace*{-3mm}\caption[]{#2} \label{#3} \
}\end{minipage}\hspace*{0.5in}\
\begin{minipage}[c]{3in}{
\includegraphics[width=3in]{#4} 
\vspace*{-3mm}\caption[]{#5}\label{#6} \
}\end{minipage}
\end{center}
\vspace*{-0.4in}\
\end{figure*}
}
}

\def\threefigure[#1,#2,#3,#4,#5]{
\begin{figure*}
\vspace*{0mm}
\begin{center}

\begin{tabular}{ccc}
\includegraphics[width=2in]{#1} & \includegraphics[width=2in]{#2} &  \includegraphics[width=2in]{#3} \\
(a) & (b) & (c) \\
\end{tabular}

\vspace*{-3mm}\caption[]{#4
} \label{#5}

\vspace*{-5mm}
\end{center}
\vspace*{-2mm}
\end{figure*}}

\def\dssfigure[#1,#2,#3,#4,#5,#6]{
{
\begin{figure*}
\vspace*{0.2in}\
\begin{center}
\begin{minipage}[c]{4in}{
\includegraphics[width=4in]{#1}
\vspace*{-3mm}\caption[]{#2} \label{#3} \
}\end{minipage}\hspace*{0.5in}\
\begin{minipage}[c]{2in}{
\includegraphics[width=2in]{#4}
\vspace*{-3mm}\caption[]{#5}\label{#6} \
}\end{minipage}
\end{center}
\vspace*{-0.4in}\
\end{figure*}
}
}

\def\dsfigure[#1,#2,#3,#4,#5,#6]{
{
\begin{figure*}
\vspace*{0.2in}\
\begin{center}
\begin{minipage}[c]{3in}{
\includegraphics[width=3in]{#1}
\vspace*{-3mm}\caption[]{#2} \label{#3} \
}\end{minipage}\hspace*{0.5in}\
\begin{minipage}[c]{3in}{
\hspace*{0.5in}\
\includegraphics[height=3in]{#4}
\vspace*{-3mm}\caption[]{#5}\label{#6} \
}\end{minipage}
\end{center}
\vspace*{-0.4in}\
\end{figure*}
}
}

\def\dsyfigure[#1,#2,#3,#4,#5,#6]{
{
\begin{figure*}
\vspace*{0.2in}\
\begin{center}
\begin{minipage}[c]{2.5in}{
\includegraphics[height=2.5in]{#1}
\vspace*{-3mm}\caption[]{#2} \label{#3} \
}\end{minipage}\hspace*{0.5in}\
\begin{minipage}[c]{2.5in}{
\includegraphics[height=2.5in]{#4}
\vspace*{-3mm}\caption[]{#5}\label{#6} \
}\end{minipage}
\end{center}
\vspace*{-0.4in}\
\end{figure*}
}
}

\def\dyfigure[#1,#2,#3,#4,#5,#6]{
{
\begin{figure*}
\vspace*{0.2in}\
\begin{center}
\begin{minipage}[c]{3in}{
\includegraphics[height=3in]{#1} 
\vspace*{-3mm}\caption[]{#2} \label{#3} \
}\end{minipage}\hspace*{0.5in}\
\begin{minipage}[c]{3in}{
\includegraphics[height=3in]{#4} 
\vspace*{-3mm}\caption[]{#5}\label{#6} \
}\end{minipage}
\end{center}
\vspace*{-0.4in}\
\end{figure*}
}
}

\def\dyoldfigure[#1,#2,#3,#4,#5,#6]{
{
\begin{figure*}
\vspace*{0.2in}\
\begin{center}
\begin{minipage}[c]{3in}{
\epsfysize=2.0in\
\hspace{0.5in}\
\epsfbox{#1}
\vspace*{-3mm}\caption[]{#2} \label{#3} \
}\end{minipage}\hspace*{0.25in}\
\begin{minipage}[c]{3in}{
\epsfysize=2.0in\
\hspace{0.5in}\
\epsfbox{#4}
\vspace*{-3mm}\caption[]{#5}\label{#6} \
}\end{minipage}
\end{center}
\vspace*{-0.4in}\
\end{figure*}
}
}

\def\cfiguredouble[#1,#2,#3,#4]{
\begin{figure}
\vspace*{0.2in}\
\begin{center}
\begin{minipage}[c]{1.5in}{
\epsfxsize=1.5in\
\epsfbox{#1}
}\end{minipage}\hspace*{0.1in}\
\begin{minipage}[c]{1.5in}{
\epsfxsize=1.5in\
\vspace{0.1in}\epsfbox{#2}
}\end{minipage}\vspace*{-0.10in} \caption[]{#3}\label{#4}
\end{center}
\vspace*{-0.4in}\
\end{figure}
}

\def\wpfigure[#1,#2,#3,#4]{
\begin{figure*}
\vspace*{4mm}
\begin{center}

\includegraphics[width=#4]{#1} 

\vspace*{-3mm}\caption[]{#2
} \label{#3}

\vspace*{-5mm}
\end{center}
\end{figure*}}

\def\wprfigure[#1,#2,#3,#4,#5]{
\begin{figure*}
\vspace*{4mm}
\begin{center}

\includegraphics[width=#4, angle=#5]{#1} 

\vspace*{-3mm}\caption[]{#2
} \label{#3}

\vspace*{-5mm}
\end{center}
\end{figure*}}

\def\DoubleFigureWSlide[#1,#2,#3,#4,#5,#6,#7,#8,#9]{
\begin{figure*}
\vspace*{#9}
\begin{center}
\begin{minipage}{#4}
\includegraphics[width=#4]{#1}
\vspace*{-3mm}\caption{#2
}\label{#3}
\end{minipage}
\hspace{2em}
\begin{minipage}{#8}
\includegraphics[width=#8]{#5}
\vspace*{-3mm}\caption{#6
}\label{#7}
\end{minipage}
\vspace*{-5mm}
\end{center}
\end{figure*}
}

\def\DoubleFigureW[#1,#2,#3,#4,#5,#6,#7,#8]{
\begin{figure*}
\vspace*{0in}
\begin{center}
\begin{minipage}{#4}
\includegraphics[width=#4]{#1}
\vspace*{-3mm}\caption{#2
}\label{#3}
\end{minipage}
\hspace{2em}
\begin{minipage}{#8}
\includegraphics[width=#8]{#5}
\vspace*{-3mm}\caption{#6
}\label{#7}
\end{minipage}
\vspace*{-5mm}
\end{center}
\end{figure*}
}

\def\DoubleFigureWHack[#1,#2,#3,#4,#5,#6,#7,#8]{
\begin{figure*}
\vspace*{0in}
\begin{center}
\begin{minipage}{3in}
\includegraphics[width=#4]{#1}
\vspace*{-3mm}\caption{#2
}\label{#3}
\end{minipage}
\hspace{2em}
\begin{minipage}{3in}
\includegraphics[width=#8]{#5}
\vspace*{-3mm}\caption{#6
}\label{#7}
\end{minipage}
\vspace*{-5mm}
\end{center}
\end{figure*}
}

\def\ddcfigure[#1,#2,#3,#4]{
\begin{figure*}
\vspace*{0.2in}\
\begin{center}
\begin{minipage}[c]{3in}{
\includegraphics[height=3in]{#1} 
}\end{minipage}\hspace*{0.5in}\
\begin{minipage}[c]{3in}{
\includegraphics[height=3in]{#2} 
}\end{minipage}\vspace*{-0.10in} \caption[]{#3}\label{#4}
\end{center}
\vspace*{-0.4in}\
\end{figure*}
}

\def\ddcfigureSlide[#1,#2,#3,#4,#5]{
\begin{figure*}
\vspace*{#5}\
\begin{center}
\begin{minipage}[c]{3in}{
\includegraphics[height=3in]{#1} 
}\end{minipage}\hspace*{0.5in}\
\begin{minipage}[c]{3in}{
\includegraphics[height=3in]{#2} 
}\end{minipage}\vspace*{-0.10in} \caption[]{#3}\label{#4}
\end{center}
\vspace*{-0.4in}\
\end{figure*}
}

\def\cxfigure[#1,#2,#3]{
\begin{figure}
\vspace*{4mm}
\begin{center}
 
\epsfxsize=2.5in\
\epsfbox{#1}\
 
\vspace*{-0.10in}\caption[]{#2
} \label{#3}
 
\vspace*{-5mm}
\end{center}
\vspace*{-2mm}
\end{figure}}

\usepackage{xcolor}
\definecolor{commentgreen}{RGB}{2,112,10}
\definecolor{eminence}{RGB}{108,48,130}
\definecolor{weborange}{RGB}{255,165,0}
\definecolor{frenchplum}{RGB}{129,20,83}

\usepackage{listings}
\usepackage{geometry}
\newcommand{\us}[1]{$\mu$s}

\newcommand{\x}[1]{$\times$}
\definecolor{pink}{rgb}{1.0,0.47,0.6}
\definecolor{orange}{rgb}{1.0,0.5,0.0}
\definecolor{cyanish}{rgb}{0,0.8,1.0}

\newcommand{\mycommand}[1]{\texttt{#1}}
\newcommand{\LogWrite}[1]{\mycommand{LogWrite}}
\newcommand{\Commit}[1]{\mycommand{Commit}}
\newcommand{\Abort}[1]{\mycommand{Abort}}
\newcommand{\WriteBack}[1]{\mycommand{WriteBack}}
\newcommand{\AtomicWrite}[1]{\mycommand{AtomicWrite}}
\newcommand{\ShortAtomicWrite}[1]{\mycommand{ShortAtomicWrite}}

\newcommand{\mystate}[1]{\textsc{#1}}
\newcommand{\Free}[1]{\mystate{Free}}
\newcommand{\Pending}[1]{\mystate{Pending}}
\newcommand{\Committed}[1]{\mystate{Committed}}

\newcommand{\Valid}[1]{\mystate{Valid}}
\newcommand{\Marked}[1]{\mystate{Marked}}

\newcommand{\eg}{\textit{e.g.}}
\newcommand{\ie}{\textit{i.e.}}

\newcommand{\MB}{~MB}
\newcommand{\GB}{~GB}
\newcommand{\TB}{~TB}

\newcommand{\gbps}{~\,Gbps}

\newcommand{\ms}{~\mbox{$ms$}}

\newcommand{\beforecaption}{\vspace{-.15cm}\begin{spacing}{0.85}}
\newcommand{\aftercaption}{\vspace{-.45cm}\end{spacing}}
\newcommand{\mycaption}[3]{\beforecaption\caption{\label{#1}{\bf \footnotesize #2} \em\footnotesize #3}\aftercaption}

\newcommand{\X}{{$\times$}}

\newcommand{\sys}{BulkX\xspace}

\newcommand{\xsplit}{\texttt{@compute}\xspace}

\newcommand{\remote}{\texttt{@data}\xspace}

\newcommand{\ufd}{\texttt{userfaultfd}}

\newcommand{\zhiyuan}[1]  {\noindent{\color{orange} {\bf \fbox{Zhiyuan}     {\it#1}}}}

\usepackage{calc}

\usepackage[textsize=tiny,textwidth=0.6in]{todonotes}
\newcommand{\allnotes}[1]{}

\newcommand{\noteyiying}[1]{\allnotes{\todo[color=green!50]{YZ: #1}}}

\newcommand{\notezhiyuan}[1]{\allnotes{\todo[color=red!50]{ZG: #1}}}

\begin{document}

\if 0
\twocolumn[\begin{@twocolumnfalse}

\begin{centering}
\vspace{0.1cm}
{\large \bf \sys: Resource-Centric Serverless for Bulky Application\\}
\vspace{0.15cm}
%OSDI'24 Submission \# 
%\vspace{-0.3cm}

\end{centering}

\bigskip

\end{@twocolumnfalse}]
\fi

\title{\sys: Resource-Centric Serverless for Bulky Application}
\date{}

\author{
 Zhiyuan Guo \qquad 
 Zachary Blanco\qquad
Zerui Wei \qquad
Junda Chen \qquad
 Jinmou Li \\
  Bili Dong \qquad
 Ishaan Pota\textsuperscript{$\mathsection$} \qquad
 Mohammad Shahrad\textsuperscript{*} \qquad
 Harry Xu\textsuperscript{$\mathsection$} \qquad
 Yiying Zhang \\ \\
 %}
 %\date{\it{
 %\textsuperscript{\textdagger}
 UC San Diego \qquad
 \textsuperscript{*}University of British Columbia \qquad
 \textsuperscript{$\mathsection$}UC Los Angeles
 }

\maketitle

%\thispagestyle{empty}

% \section*{Abstract}
\begin{abstract}

Serverless computing, commonly offered as Function-as-a-Service, was initially designed for small, lean applications. However, there has been an increasing desire to run larger, more complex applications (what we call {\em bulky} applications) in a serverless manner. Function-based serverless systems cause significant resource wastage when executing bulky applications, as a function is the resource allocation and execution unit, yet function size does not capture application needs.

We propose a new resource-centric serverless-computing model for executing bulky applications, and we build the \sys\ serverless platform following this model.
The core idea of \sys\ is {\em adaptive serverless}, where resource allocation and execution automatically and dynamically adapt to application behavior and underlying cluster resource availability.
%in a resource- and performance-efficient way
Our results show that \sys\ reduces resource consumption by up to 90\% compared to today's function-based serverless systems,
%\notemohammad{suggestion: "today's function-centric resource allocation model" instead of "today's FaaS"} 
while %adding only 1.3\% performance overhead for some application and 
improving performance by up to 64\%. % for others.
%solves various resource-related issues of today's serverless computing, while retaining or even  improving its performance.

\end{abstract}
\section{Introduction}
\label{sec:intro}

Serverless computing, commonly offered as Function-as-a-Service (FaaS), is a cloud service that allows users to deploy and execute their applications without managing servers.
Serverless computing has gained tremendous popularity in the past few years thanks to its benefit of minimal IT burdens, pay-per-use pricing, and automatic scaling (auto-scaling)~\cite{serverless-growth-1,serverless-growth-2}.

Initially, serverless computing was designed for running short, small-scale functions like simple HTML serving. 
In recent years, there has been an increasing desire to run large, complex applications such as video processing~\cite{excamera,ao2018sprocket},
scientific computing~\cite{numpywren,serverlessseqcompar19}, machine-learning tasks~\cite{carreira2019cirrus,AWS-SageMaker-Serverless,Azure-ML-Training-Serverless}, data analytics~\cite{Databricks-SQL-Serverless,AWS-Redshift-Serverless}, and relational databases~\cite{AWS-Aurora-Serverless,Azure-SQL-Serverless-Announce} in a serverless style (\ie, event-triggered, pay-per-use pricing, auto-scaling, and no management of underlying systems). These applications often 1) run longer or consume more memory than typical FaaS function size limits, 2) exhibit resource usage variations across different phases of computation, 3) require different amounts of resources with different inputs.
We call serverless applications with one or more of these features {\em bulky serverless applications}, or bulky applications for short.

Today's practice of running bulky applications in a serverless style is function DAGs~\cite{AWS-Step-Functions,jonas2017occupy_pywren,excamera,azure_durable_functions,OpenWhisk-composer,zhang2020kappa}, where users construct a DAG of functions and set the size of each function in the DAG. This approach creates several resource-waste and performance issues.
First, a function's size is fixed across invocations. Yet, under different inputs,
%to a bulky application often have different resource requirements~\cite{bilal2021,mvondo2021ofc}, resulting in 
there will be resource waste if provisioning for peak input usage or application failure if under-provisioning.
Second, different functions in a DAG execute in different environments (\eg, containers on one server or different servers). \noteyiying{need ref for this}
Communicating between and setting up the environments can become a major performance overhead, especially when breaking a bulky application into many functions.
On the other hand, reducing the number of environments by making functions bigger would cause resource waste, as a function has the same resource allocation throughout its execution; resources allocated for the peak execution needs are wasted during non-peak time of the function.
Finally, functions in a DAG store intermediate and shared data in a disaggregated layer, usually over a key-value-like interface~\cite{klimovic2018pocket,AWS-Step-Functions}.
Accessing such a layer causes performance overhead due to network communication and data serialization.
The root cause of these issues is today's \textit{function-centric} serverless model, where the resource size, resource unit (code piece or data objects), and execution method of a function cannot adapt to application workloads or hardware resource availability.
%\notezhiyuan{term: code scope}

{
\begin{figure*}[t]
\begin{center}
%\centerline{
\includegraphics[width=\textwidth]{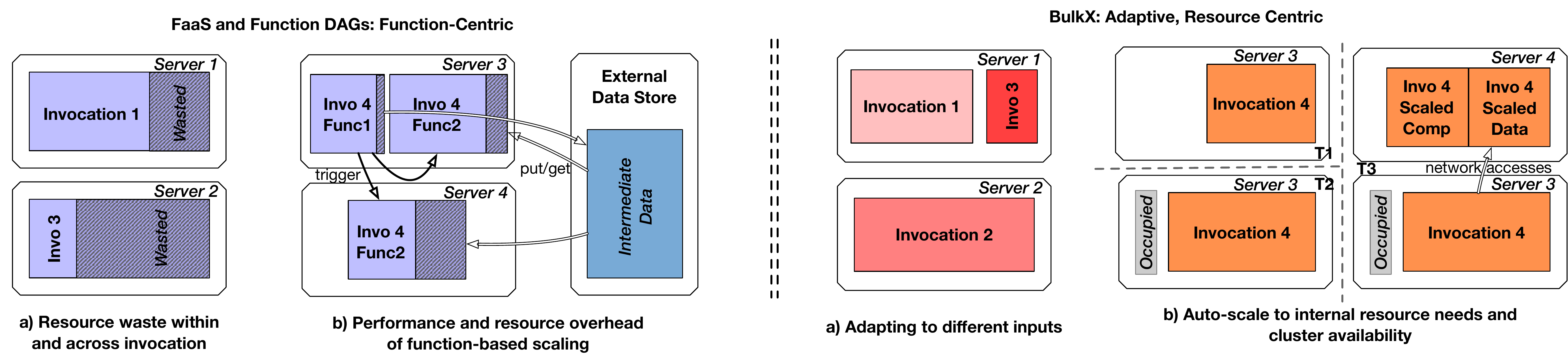}
%}
\vspace{-0.2in}
\mycaption{fig-faasraas}{Function- vs. Resource-Based Serverless Computing.} 
{
 The left half shows today's FaaS and function DAGs, which incur significant resource waste and communication performance overhead.
 The right half shows how \sys\ adapts resource allocation and execution across invocations.
}
\end{center}
\vspace{-0.1in}
\end{figure*}

{
\begin{figure}[t]
\begin{center}
%\centerline{
\includegraphics[width=\columnwidth]{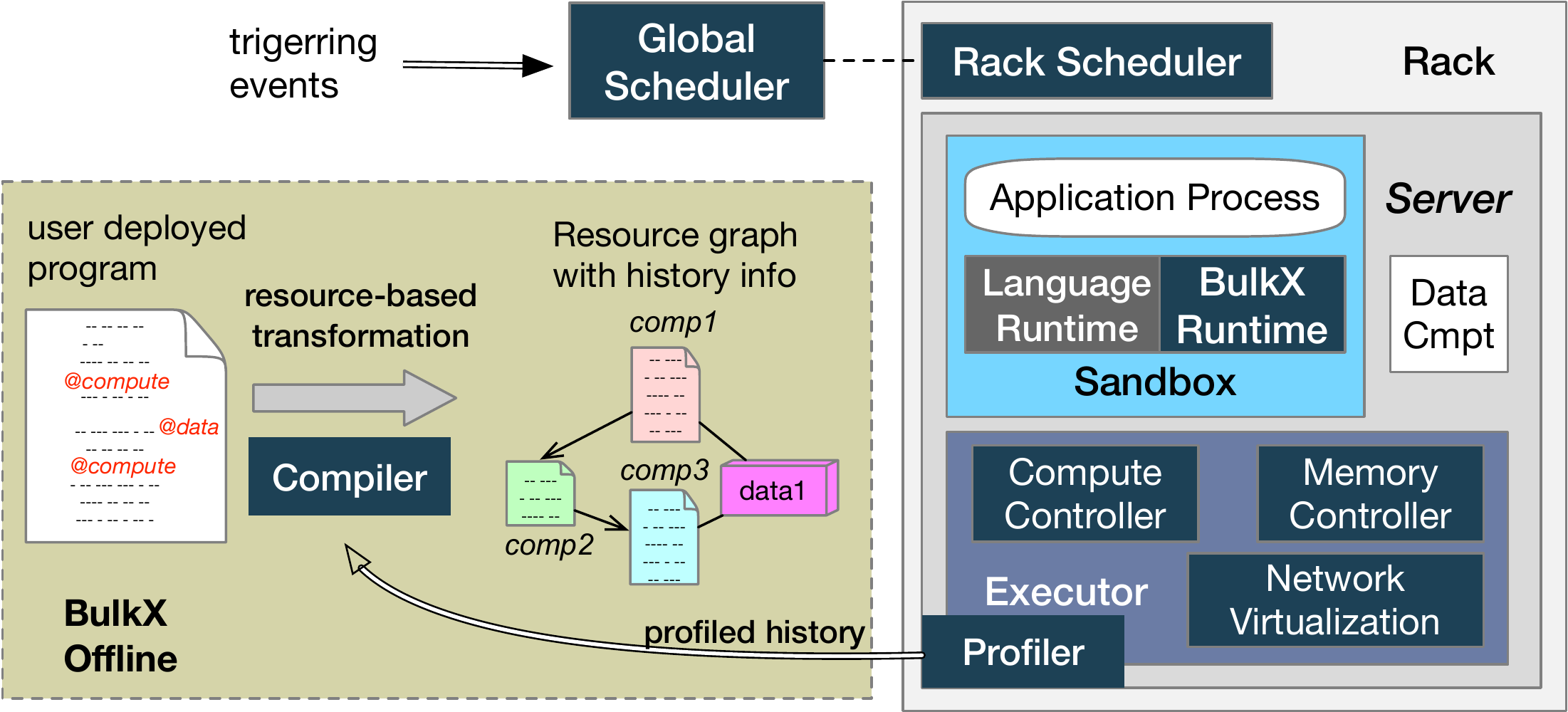}
%}
\vspace{-0.1in}
\mycaption{fig-overall}{Overall Workflow and \sys\ Components.}
{
The \sys\ offline part runs behind the scene. The rest runs when user applications are triggered.
}
\end{center}
\vspace{-0.1in}
\end{figure}

To address these issues, we propose an \textbf{\textit{adaptive, resource-centric}} serverless model:
%The main idea behind \sys\ is \textbf{\textit{adaptive serverless}}, 
users deploy monolithic programs; at each invocation, the serverless system adapts the resource placement, scaling, and execution methods to tightly follow the {\em internal} resource requirements of that invocation and the current cluster resource availability (Figure~\ref{fig-faasraas}).
%\notezhiyuan{detailed approach?}
%For example, an application can execute in its entirety in one container process if a server can meet all its resource needs throughout one invocation. For another invocation, it can be executed as two containers on two servers accessing memory on a third server followed by ten parallel executions of a task on five servers. Such adaptation is fully automated with the goal of minimizing resource waste and performance overhead. \noteyiying{last sentence can be improved}
%\notezhiyuan{Auto scale at runtime}

Following this model, we build \textbf{\textit{\sys}}, a serverless platform designed for bulky applications. %Users submit monolithic programs to \sys\ instead of function DAGs.
%\sys\ decides the allocation and execution units based on an application's resource %needs; the unit can be a whole application or a part of it. For each unit, 
To be adaptive, \sys\ supports several execution methods and scopes. 
First, \sys\ supports the execution of a part of or the whole application in one environment, where all computation and memory accesses are native, \ie, unmodified memory instructions and function calls within a process (\eg, invocations 1 and 2 in Figure~\ref{fig-faasraas}).
Then, for memory usage, in addition to local memory, \sys\ supports the placement of memory objects on a different server from computation and transparently converts applications' memory accesses into network communication (\eg, invocation 4 T3 data in Figure~\ref{fig-faasraas}). For computation, \sys\ supports the parallel execution of a code piece on one or multiple servers and invoking a function in the user program in a local process or on a remote server. 
%\notezhiyuan{term object}
%may schedule different parts of an application's computation as parallel instances or subsequent invocations on one or multiple servers, and it automatically determines the total CPU cores to use. Depending on their locations, different computation parts could communicate as normal function calls or as network calls.

To execute applications using the above mechanisms, \sys\ should first choose the appropriate resource unit, resource size, and resource location to meet an application invocation's resource needs (\ie, resource-centric).
To accomplish this \sys\ must capture resource features and their changes within an application's invocation and across invocations.
Instead of imposing runtime monitoring overhead and reacting to every resource change in an on-demand way, our solution is to capture as many resource features prior to the execution as possible, with a combination of user hints and coarse-grained profiling history, so that at runtime, \sys\ can make {\em proactive} scheduling decisions and sets up execution and communication environments in the background.
Specifically, users express where resource changes may happen by annotating their source programs.
\sys\ identifies the smallest execution units based on user-annotated {\em code scopes} and transforms user programs into an intermediate representation we proposed called {\em resource graph}. Each node in the graph corresponds to a user annotation, as shown in Figure~\ref{fig-overall}. 
To decide the size (number of cores, amount of memory) of each graph node, \sys\ leverages profiled history when making on-demand, auto-scaling decisions.
%\noteyiying{this part is too complex to explain in intro. this is the best i can do. i think it's fine. no one expects to understand the whole thing by just reading the intro}
%resource amounts and lifetime in an application's past invocations. At runtime, \sys\ automatically scales resource amounts uses this history
%At runtime, \sys\ makes resource scheduling decisions based on an application's resource graph and cluster resource availability and {\em proactively} sets up execution and network environments in the background before they are needed. To handle new resource needs not captured in the history, \sys\ makes on-demand, post-allocation resource adjustments. 
%Compared to a pure on-demand approach, \sys\ largely avoids the performance overhead of execution- and network-environment setup.
To place a graph node, \sys\ uses a locality-based, greedy placement policy. 
\sys\ co-locates computation and memory as much as possible, first in one server, multiple servers within a rack, and finally across racks. 

In sum, unlike serverless DAGs, \sys's internal resource graph and profiled history capture fine-grained resource features of an application, and \sys\ executes adaptively by ``merging'' graph nodes into local execution, by supporting non-local execution, and by automatically sizing each execution unit. As such, \sys\ achieves optimized performance and resource efficiency for each application invocation.

We evaluate \sys\ with a set of microbenchmarks and three real-world applications: data analytics queries from the TPC-DS benchmark, video processing pipelines, and a machine-learning task,
We compared \sys\ to two FaaS frameworks (OpenWhisk~\cite{OpenWhisk}, AWS Lambda~\cite{AWS-Step-Functions}), four serverless DAG frameworks (AWS Step Functions~\cite{AWS-Step-Functions}, PyWren~\cite{jonas2017occupy_pywren}, ExCamera~\cite{excamera}, gg~\cite{fouladi2019laptop}), one serverless resource tuning system (Orion~\cite{Orion-osdi22}), and two VM-based execution methods (remote-memory swapping~\cite{FastSwap}, VM migration~\cite{migros}).
Our machine-learning results show that \sys\ reduces resource consumption by 40\% to 84\% compared to OpenWhisk while only adding 1.3\% performance overhead.
Our TPC-DS results show that \sys\ reduces resource consumption by 73\% to 85\% and improves performance by 54\% to 64\% compared to PyWren.
Our video process results show that \sys\ reduces resource consumption by 33\% to 90\% and improves performance by 33\% to 47\% compared to gg on Openwhisk.

We will make \sys\ publicly available upon acceptance.
         % section 1
% \input{intro-zhiyuan2}
%\input{intro-2}         % section 2
%\input{introduction}    % section 3
\section{Motivation and Related Works}
\label{sec:background}

This section discusses serverless application trends, the limitations of function-centric serverless computing, and various related works.

\begin{figure}[t]
\begin{minipage}{0.48\columnwidth}
\begin{center}
\centerline{\includegraphics[width=\columnwidth]{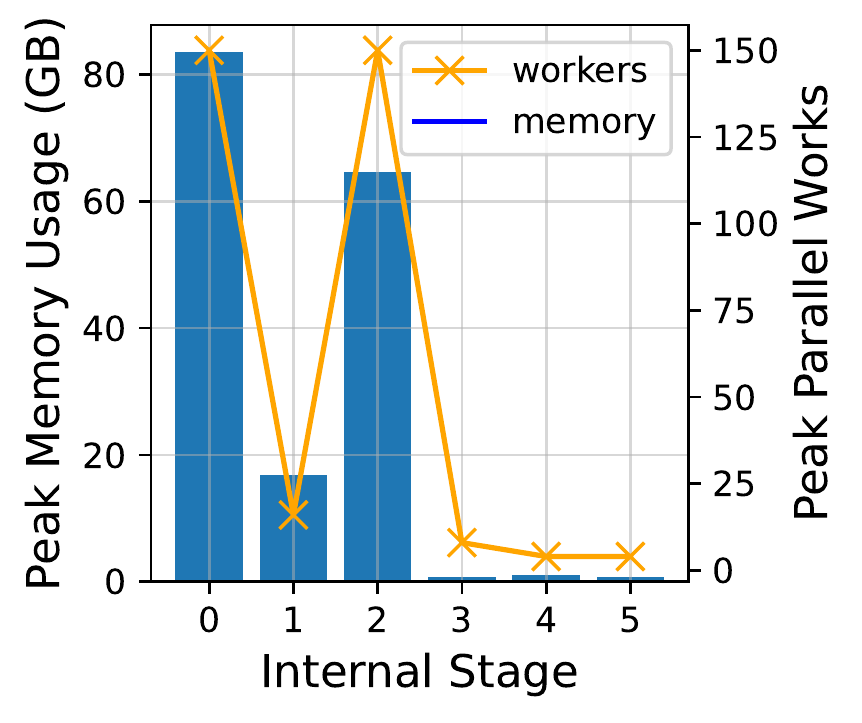}}
\vspace{-0.15in}
\mycaption{fig-object}{Resource Variation within one Invocation.}
{
Running the TPC-DS Query 95 with 100\GB\ data as input.
}
\end{center}
\end{minipage}
\hspace{0.005\columnwidth}
\begin{minipage}{0.48\columnwidth}
\begin{center}
\centerline{\includegraphics[width=\columnwidth]{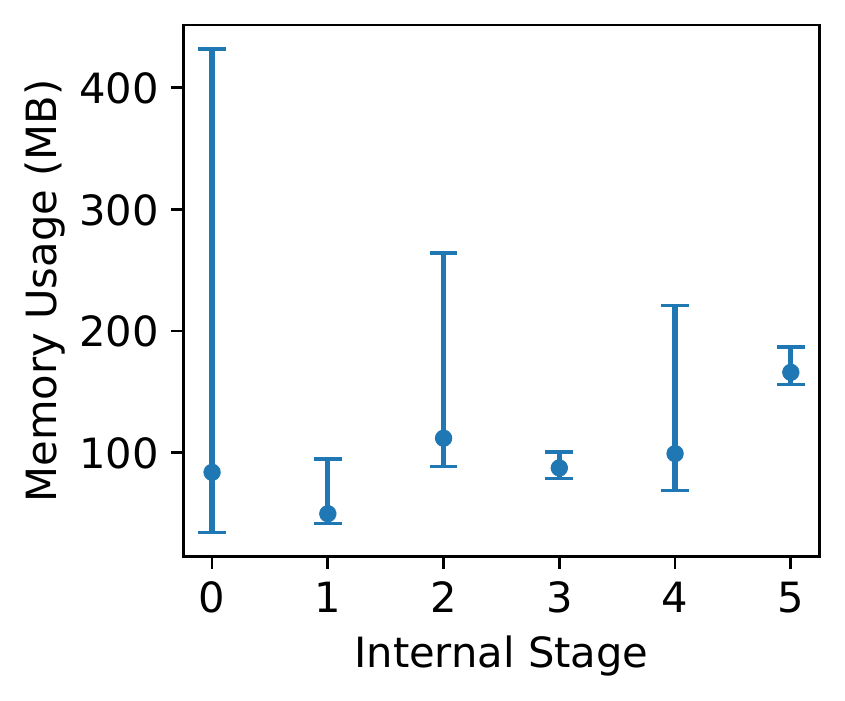}}
\vspace{-0.15in}
\mycaption{fig-invocation}{Resource Variations across Inputs.}
{Running TPC-DS Query 95. Marks on lines show min, max, and avg values across inputs.}
\end{center}
\end{minipage}
% \begin{minipage}{0.48\columnwidth}
% \begin{center}
% \centerline{\includegraphics[width=\columnwidth]{Figures/Figure-motiv-lr.pdf}}
% \vspace{-0.15in}
% \mycaption{fig-invocation}{Resource usage with different inputs.}
% {Running MP4 add watermark~\cite{copik2021sebs}.
% }
% \end{center}
% \end{minipage}
\if 0
\hspace{0.005\columnwidth}
\begin{minipage}{0.68\columnwidth}
\begin{center}
\centerline{\includegraphics[width=\columnwidth]{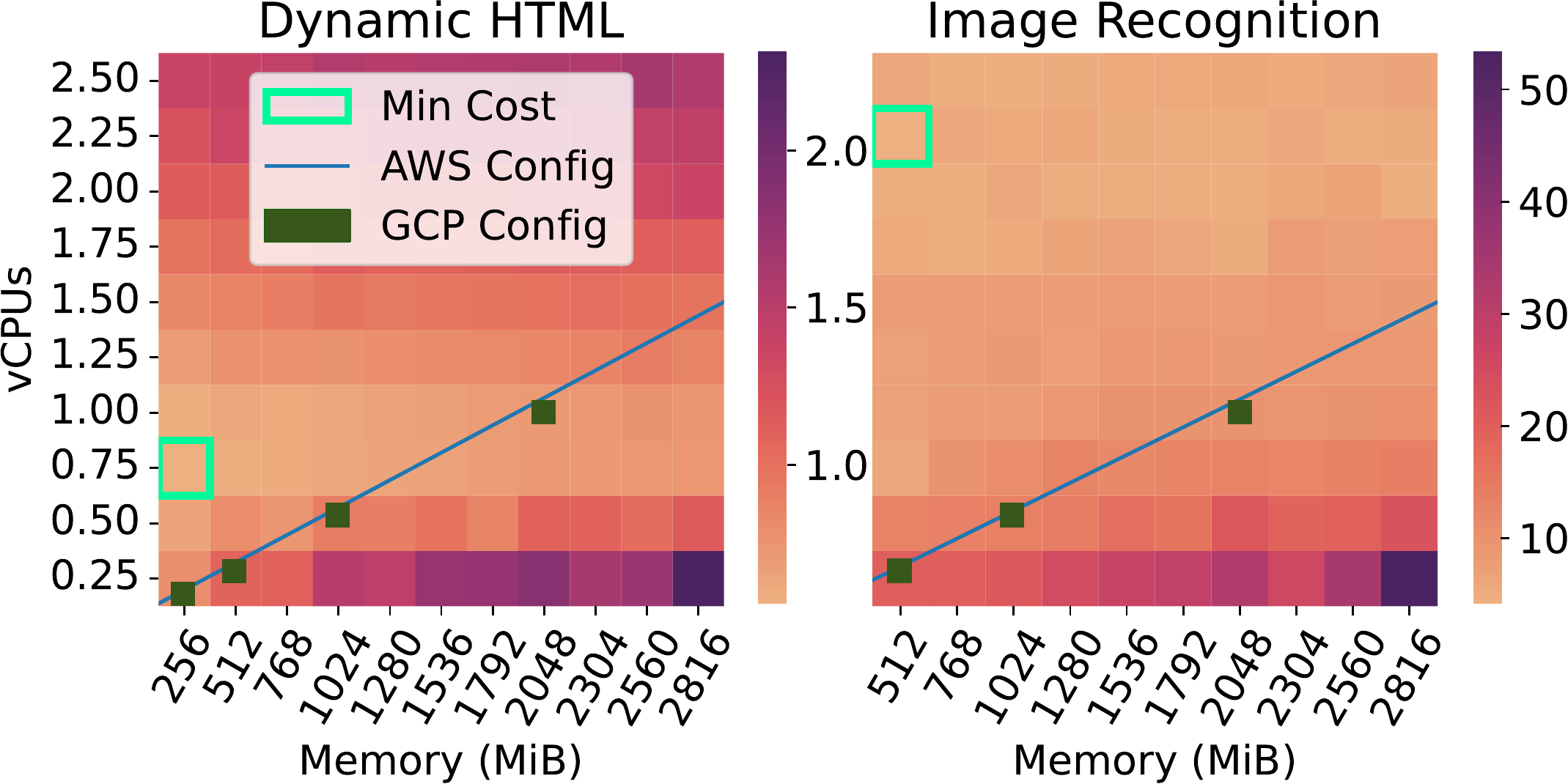}}
% \vspace{-0.1in}
\mycaption{fig-heatmap}{Resource Consumption Heatmap.}
{
Each cell runs in a container on a local server with memory and vCPU sizes configured as X and Y axis values. 
Lighter color is better (smaller cost).
}
\end{center}
\end{minipage}
\hspace{0.005\columnwidth}
\begin{minipage}{0.65\columnwidth}
\begin{center}
\centerline{\includegraphics[width=\columnwidth]{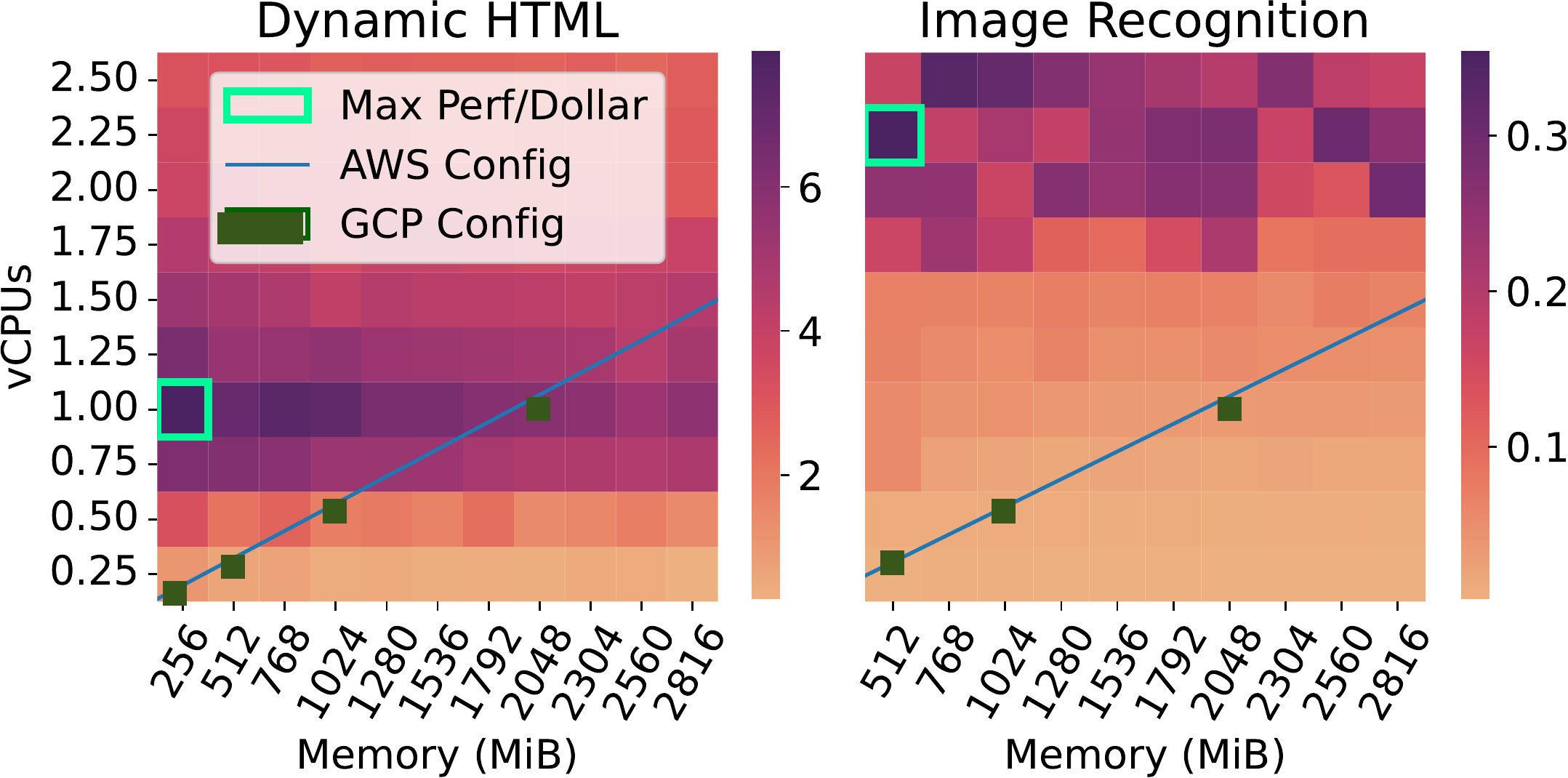}}
% \vspace{-0.1in}
\mycaption{fig-heatmap-perfdollar}{Perf-per-Dollar Heatmap.}
{
Performance per dollar calculated using workload execution time and cost as seen in Figure 1.
Darker color is better (higher performance per dollar).
}
\end{center}
\end{minipage}
\fi
%\vspace{-0.1in}
\end{figure}

%\notezhiyuan{could consdier 2.0. Bulky application need resource flixibiility}
\subsection{Serverless Application Trends}
% \fixme{To be filled by Mohammad}

% application trend (complexity, size, workflows, languages, richness, etc.)
Serverless applications are becoming increasingly more complex and bulkier.
According to a 2021 study, 31\% of open-source serverless applications have workflow structures~\cite{eismann2021state}.
The popularity of serverless DAGs at Azure witnessed a 6$\times$ increase from 2019 to 2022~\cite{Mahgoub_wisefuse_2022}.
Meanwhile, large applications that traditionally run in VMs, such as databases and machine learning, are seeing serverless offerings  with better resource and cost efficiency~\cite{Azure-SQL-Serverless-Announce,Databricks-SQL-Serverless,AWS-SageMaker-Serverless,pu2019shuffling}.
%; they consume more resources or run longer than traditional serverless applications.
%This increasing complexity has translated into building richer application classes on top of serverless. %As another evidence, serverless offerings are adding the support of languages usually used for writing more complex, larger applications like Java, which is now the third most popular serverless language~\cite{datadog23}.

As applications get more complex and larger, their executions internally have different resource requirements.
For example, as shown in Figure~\ref{fig-object}, the TPC-DS~\cite{tpcds:sigmod02} data-analytics benchmark's query 95 includes five internal stages. They demand drastically different amounts of memory and CPU (as seen by the number of parallel workers). 
%As another example, we measure the resource demands of a logistic-regression-based training application~\cite{carreira2019cirrus}. Its highest and lowest memory usage phases differ by 15.5 times.
Other applications, such as machine learning, video/image processing workflows, and scientific computing, also exhibit resource-consumption variations within one execution.
%, and efforts have been made to add serverless support for each of these domains.
%\notezhiyuan{generally we should not use computation phases - as the structure is not a linear porgram with different phases. thinking about a new term}

%\zhiyuan{Inside one non-speratable space},
At the same time, many serverless applications' resource consumption depends on their inputs~\cite{mvondo2021ofc,Eismann_2021,Moghimi_socc_23}.
%\notemohammad{disagree with the point above. how about we write it this way: As applications get bulkier, the same proportion of input-dependent resource variations translates into a greater amount of resource wastage.}
Figure~\ref{fig-invocation} shows the memory usages of the same TPC-DS query 95 when given input sizes ranging from 10\GB\ to 200\GB{}. %, which downloads an MP4 video file and adds a watermark to it. %The second stage exports this new MP4 to the GIF format.
Across different inputs, the variation of memory usage at each stage can be as huge as 12 times for the same worker function.
%Even under one input, different parallel worker in the same execution.
We observe similar input-dependent resource features in other data analytics (33\x{} more resource when the TCP-DS benchmark dataset size grows by 10\x{}), video transcoding (94\x\ differences when input video changes from 240P to 4K), and many others.
%, and machine learning training (3\x\ differences when input size grows from 12\MB\ to 44\MB{}). 
%Others also report similar input-dependent features of serverless applications

%Today's serverless platforms also offer better support for more resource-intensive language runtimes. 
%A recent report by Datadog, analyzing more than 20,000 serverless telemetry customers, states that Java is now the third most popular serverless language \textit{``likely due to a growing number of large enterprise organizations migrating their existing workloads and applications written in Java over to Lambda.''}~\cite{datadog23}.

% Recent (Nov. 2023) serverless characterization data from a public serverless provider reports low CPU and memory utilization~\cite{Joosen_socc_23}.
% They state \textit{``60\% of all allocations using less than 0.1 cores but asking for a limit of more than 1 core''} and \textit{``around 50\% of the functions having a limit of around 2 GB, but using only around 400 MB''}.
% Prior work has attributed this underutilization primarily to users setting resource limits conservatively to prevent function failure~\cite{Joosen_socc_23,Moghimi_socc_23}.
% In particular, the requirement to set static limits 

% in face of 1) input-dependent resource variations~\cite{Moghimi_socc_23,mvondo2021ofc}, and to match the peak resource demand during the execution. 

%\fixme{Mohammad still working on this}

% more applications that have input-dependent resource consumption

% applications with varying resource consumption phases

\subsection{Function-Based Serverless Limitations}
\label{sec:serverless}

Existing serverless computing solutions mainly follow a function-based model, running applications as a single function or a DAG of functions. We now discuss their limitations for bulky applications.
%function model, where applications execute as functions with a particular size (CPU, memory) chosen by the user, an automated tool, or the provider.

\boldunderpara{Resource waste in functions.}
A Nov 2023 serverless characterization study from a public serverless provider reports more than 90\% CPU wastage for 60\% of invocations and 80\% memory waste for around 50\% of functions~\cite{Joosen_socc_23}.
%\textit{``60\% of all allocations using less than 0.1 cores but asking for a limit of more than 1 core''} and \textit{``around 50\% of the functions having a limit of around 2 GB, but using only around 400 MB''}~\cite{Joosen_socc_23}.
The resource inefficiency of today's function-centric serverless computing stems from its fixed function size throughout the execution of a function invocation and across different invocations. Moreover, today's FaaS providers like AWS Lambda only support fixed, limited sets of function CPU-to-memory ratios~\cite{google_functions_pricing,aws_lambda_config_mem}, but workloads can have arbitrary CPU-to-memory ratios~\cite{bilal2021}.
%, causing CPU or memory wastage even with optimized selection of function sizes~\cite{power-tuning,Eismann_2021,COSE,ryan-serverless}.  These resource inefficiencies become worse with bulky applications.

%\boldunderpara{Function resource tuning.}
To help users save cost, research works~\cite{Orion-osdi22,ryan-serverless} and tools~\cite{power-tuning,Eismann_2021,COSE} find cost-optimal function sizes in public clouds. %A recent work, Orion~\cite{Orion-osdi22}, improves the prior state of the art by using execution time distribution history to optimize function sizes. %but Orion's selected size is still the same for all invocations of the function.
These works are still confined by the function model and all its limitations. For example, Orion selects an optimal size for a function but has to use the same size throughout the function's entire execution and for all invocations of the function. Moreover, the most optimal CPU-to-memory ratios may not fall into the few options offered by providers. \sys\ avoids such limitations altogether by adapting resources according to applications' execution needs.

%\zhiyuan{Performance?}
\boldunderpara{Function DAGs.}
Several serverless frameworks like Durable Functions~\cite{burckhardt2021serverless}, AWS Step Function~\cite{AWS-Step-Functions}, and OpenWhisk Composer~\cite{OpenWhisk-composer} let users express their applications as a DAG of functions. 
%It further provides interfaces for programmers to mark the persistence point of their application to enable fast failure recovery. \sys\ uses a similar recovery philosophy but does so automatically without introducing new programming models. More important, Durable Functions (and other FaaS chaining frameworks like ) rely on users to provide the DAG of functions. \sys\ generates the DAG of elements based on resource usage patterns.
Kappa~\cite{zhang2020kappa} is a continuation-based programming model for long-running applications to work beyond function timeouts by splitting programs into fine-grained pieces based on user-specified continuation. %Unlike \sys, this split mainly focus on preserving states in user functions, and is not aware of different resource usages in different user functions.
These systems' main goal is to enable applications to run beyond a single function's size limits; they execute applications according to the statically written DAGs regardless of runtime resource needs or availability.
Once deployed, all invocations of an application use the same DAG and the same function sizes, and each function is executed in a separate environment.
As a result, it is difficult, if not impossible, for users to find the right DAG and function sizes: a small DAG with coarse-grained functions wastes resources for some invocations, while a complex DAG with many small functions causes environment communication and start-up performance overhead.
\noteyiying{this part could be improved by writing according to the new example in sec2.1}
%\notezhiyuan{we should directly see we are dynamic execution}

Differently, \sys\ dynamically, adaptively, and automatically optimizes application execution based on resource features and cluster availability for each invocation, resulting in optimal performance and resource utilization.
%Moreover, each function in a function DAG still suffers from resource waste and resource limits. \sys\ is eliminated from these issues, as it can execute an application locally and in a distributed, disaggregated manner.

\boldunderpara{Domain-specific serverless.}
Apart from generic serverless frameworks, there has been a body of domain-specific works for enabling various application domains to run in the serverless style~\cite{carreira2019cirrus,jonas2017occupy_pywren,excamera,zhang2021caerus,fouladi2019laptop,ao2018sprocket,perron2020starling,AWS-Redshift-Serverless,AWS-Aurora-Serverless,Azure-ML-Training-Serverless,Azure-SQL-Serverless-Announce,AWS-SageMaker-Serverless}.
For example, PyWren~\cite{jonas2017occupy_pywren} is a service that offers Python map primitives on functions that powers serverless data analytics~\cite{zhang2021caerus,Databricks-SQL-Serverless} and linear algebra~\cite{shankar2020serverless}.
%\emph{gg}~\cite{fouladi2019laptop} is a platform for building burst parallel applications and mapping them to serverless functions.
ExCamera~\cite{excamera} and Sprocket~\cite{ao2018sprocket} are serverless frameworks for video processing.
%Starling~\cite{perron2020starling} is a query execution engine running on functions. 
These frameworks all follow the function-based serverless model and inherit all the function-based resource-waste issues. Moreover, building them requires significant manual efforts for each application domain. 
Differently, \sys\ is a generic solution that avoids resource waste caused by function-based serverless.

Besides research works, there are also commercial domain-specific serverless offerings~\cite{Azure-SQL-Serverless-Announce,AWS-Aurora-Serverless,AWS-Redshift-Serverless,Databricks-SQL-Serverless,Azure-ML-Training-Serverless,AWS-SageMaker-Serverless}.
Unfortunately, no public information about their internals is available. From their pricing information, many are still following the function model. Thus, we suspect they also inherit the resource waste issues of function-based serverless.

\subsection{Resource Disaggregation and Migration}
\label{sec:disagg}

%\notezhiyuan{We could go one step back. Instead of "serverless" let's its say it enables resource flexibility}
We now discuss two techniques for reducing resource waste used in non-serverless settings, resource disaggregation and execution migration, and why they are insufficient for bulky serverless computing.
%enable bulky applications on serverless computing. As far as we know, they have not been applied to serverless yet. \noteyiying{verify this}
%We discuss how \sys\ compares with resource disaggregation and intermediate-storage serverless computing works.%, as in Table~\ref{tbl-related}.

\boldunderpara{Resource disaggregation.}
Over the past few years, there have been a host of research works to disaggregate memory from computation for resource saving~\cite{Shan18-OSDI,AIFM,Tsai20-ATC,InfiniSwap,FastSwap,RAMCloud,Pond,Mira}.
Running bulky serverless applications on a disaggregated system seems a feasible approach for resource efficiency, but this approach is insufficient.
First, unlike \sys, existing disaggregation systems do not autoscale memory or compute resources for different application invocations or execution phases, resulting in resource wastage. %Dynamic memory sizing can solve the problem, but it is hard to implement without performance overhead or scalability limitations. 
%Another potential way is to allocate at small granularity, which avoids migration but limits the scalability of the scheduling/allocation system.
%LegoOS' memory management allocates virtual memory address range at 1\GB\ chunks, but as virtual memory is freed by applications in an arbitrary way, 
%Dynamic memory sizing could solve this problem, but it is not easy to achieve dynamic sizing without incurring performance overhead or scalability limitations. 
%For example, allocated and de-allocating each object on its own would save memory but would lead to huge scheduling burden.
%, there will be a lot less memory wastage. But doing so would add huge burden to the scheduler, limiting the system's scalability.
%Furthermore, existing systems lack autoscaling support for compute pools.
%\sys\ addresses these issues by decomposing applications in a resource-centric way (thus never having the need for migration), by materializing execution units with fixed-size physical components (reducing the amount of allocations), and with a scalable, hierarchical scheduler design.
Second, disaggregation systems assume an architecture where compute nodes have insufficient memory and always have to make remote data accesses for running applications, either explicitly~\cite{AIFM,Mira} or via swapping~\cite{FastSwap,InfiniSwap}. \sys\ sits on a regular server cluster and executes an application fully locally as much as possible.
%locality and aggregation
%Finally, existing disaggregation systems focus on optimizing data-path performance, but control-path performance is critical in a serverless setting where functions (or components in \sys) can be frequently invoked. We design our scheduler and network stacks to minimize or hide control-path overhead.

%Finally, the failure handling mechanisms in today's disaggregation systems~\cite{Shan18-OSDI,XXX} are not optimized for serverless computing, which allows for invocations to be restarted with less overhead than in traditional computing.

\boldunderpara{Migration-based approaches.}
One potential way for adapting to applications' internal resource phase changes is to migrate a running container~\cite{CRIU} or other execution units~\cite{nelson2005fast,vm-live-migration-nsdi05,zap-vm-migration-osdi02,XvMotion-ATC14} to another server when the current server runs out of resources. However, live migration of a bulky application is slow, as it usually involves moving large amounts of memory. Moreover, when an application's resources grow frequently, a migration-based solution would need to either migrate repeatedly or migrate once but set the new allocation to accommodate the potential peak resource usage, causing significant performance overhead or resource wastage.
%\noteyiying{would be good to find some references that talk about high migration overhead of VM/container.}
%\notezhiyuan{Another drawback could be resource is not continues in migration: we cannot increase by a 64MB size in migration as its too slow. we need to estimate the final usage, e.g. migratie from 1G to 100G.}
Nu~\cite{Nu} proposes to decompose a process into smaller proclets, each having a non-shared heap space, while shared data is read-only and replicated across servers. The migration of proclets is lightweight, but Nu has two major drawbacks if applied to serverless computing: 1) by replicating shared data on all servers, Nu causes significant resource wastage; 2) proclets cannot write to shared data, which limits the type of applications Nu can support.
%\noteyiying{we don't support writes to shared data either, right?}
%\notezhiyuan{We can support. no coherent. but user can use sync primitives / partition accessed range}

\if 0
why resource disaggregation is not enough for serverless environments?
1. no memory management in the memory pool that gears towards efficient resource efficiency AND scheduling scalability (no autoscaling of mem pool). currently, just assumes that there's a fixed mem pool that's big enough, no deallocation => waste resource. on the other hand, if one would do allocation per access (or per object?), then the scheduling overhead would be too high
2. there's no optimization on the control paths: connection set up and execution env set up would have high overhead
3. there's no compute side scaling
4. there's no proper failure handling or too much failure handling 

The high-level idea of disaggregating data from compute has been taken by many systems~\cite{Shan18-OSDI,AIFM,Tsai20-ATC,HP-TheMachine,RAMCloud}. In fact, serverless functions like Lambda and storage layers like S3 and in-memory storage are a form of
disaggregation.
% disaggregating data from compute.
\fi

%\input{fig-workflow}
\section{\sys\ Overview}
\label{sec:overview}

To use \sys, users deploy source programs with annotations and their triggering events. Internally, \sys\ transforms the original code and dynamically chooses the most appropriate way of executing each invocation of a user application.
Centering around our idea of adaptive resource-centric serverless, \sys\ consists of several vital components in two parts: background offline tasks happening behind the scene and a foreground runtime system, as shown in Figure~\ref{fig-overall}. Connecting the two parts is an intermediate representation we proposed called {\em resource graph}. A resource graph consists of {\em virtual} computation or data resource {\em components}, each with a history-based resource profile. 

%To avoid the above problems, \sys's approach is to {\em proactively} allocate resources based on application semantics acquired offline, \ie, {\em proactive adaptive serverless}. 
\sys's offline part (\S\ref{sec:offline}) consists of a sample-based profiler, a program analyzer, and a compiler. They work together to capture resource features of an application before the \sys\ runtime carries out an invocation of it. \sys\ analyzes user-annotated programs and decomposes each application into a resource graph. Our compiler generates two versions of transferred programs, one as native execution and one as remote accesses for all the accesses across resource graph nodes. \sys\ samples application runs to perform lightweight profiling. \sys\ adds profiled historical information such as object lifetime and size to resource graphs and uses the information at the runtime for proactive, informed scheduling. 
%The \sys\ runtime system leverages node relationships and resource profiles in a resource graph to proactively allocate resources; it then chooses the appropriate compilation for execution. 
%\notezhiyuan{How the info is used?}

The foreground system, \sys\ runtime (\S\ref{sec:online}), sits at the application-perceived critical paths. Its goals are to automatically and adaptively allocate and scale CPU and memory resources and to execute applications with high resource and performance efficiency. It leverages the resource graph and the accompanying history resource profiles to {\em proactively} perform informed scheduling and scaling decisions. We design a two-level {\em scheduling system} that uses a locality-based policy to schedule as many components together or close to each other as possible. 
On each server, a \sys\ {\em executor} launches and facilitates the execution of compute and data components in containers. Within each container, application runs with the \sys\ {\em runtime library}, which executes \sys\ internal APIs such as remote memory access.
%\zhiyuan{Connect to impl structures}
%\zhiyuan{In \S\ref{sec:online}, we will focus on how we solve two main challanges: First, how to be adaptive and support multiple execution methods (\S\ref{sec:adaptive}, Second, how to reduce the on-demand scaling overhead by being proactive (\S\ref{sec:proactive}).}

%Adaptive serverless can be realized in a purely reactive, on-demand way, \eg, by moving some of an application's computation and memory to a separate server when the local server runs out of resources. Reactive adaptation has several issues in a bulky serverless environment: 1) a system would need to frequently react to resource changes to tightly pack resources, 2) adaptations can be costly when a server cannot sustain the growth of computation or data, and the increase needs to be handled by a different server, 3) resources would be fragmented due to multiple reactive adjustments.

\section{Application Deployment and \sys Offline}
\label{sec:offline}

This section introduces how to use \sys\ and discusses how \sys\ prepares efficient execution prior to the runtime, which we will introduce in \S\ref{sec:online}.

\subsection{Application Deployment}

\sys\ targets applications that are bulky, \ie, resource-hunger or long-running applications that have internal resource variations or input-dependent resource variations. Users write monolithic programs instead of functions in \sys. Unlike serverless DAGs which uses provider APIs to explicitly communicate across functions and to external data stores~\cite{numpywren,AWS-Step-Functions,azure_durable_functions}, user programs for \sys\ perform normal procedure calls within the same program as defined by the programming languages, and they use memory with native, unmodified memory allocation and accesses.
%\zhiyuan{\sys do not differentiate the access to resources is local or remote, which not only ease the effort of programming and porting, but also enables code transformation for adaptive scheduling and execution.
%Today's FaaS platforms~\cite{klimovic2018pocket,more-serverless-state-works-here} use coarse grained, static DAGs and special data-store and function invocation/ These architectures are ill-suited for the flexibility necessary in adaptive serverless.}

On top of native programs, \sys\ offers two annotations for users to identify parts of their programs that have distinctive CPU needs and memory resource needs. 
% The first annotation,  \textit{\xsplit\texttt{(max\_par=K)}}, applied to procedure calls, indicates a call site that is likely to have different parallelism from the caller, \eg, the caller function executed in a single thread, but the callee may execute in multiple parallel threads. To control their maximum budget, users can set \texttt{(max\_par=K)} to indicate that the code site's parallelism should not exceed $K$. The second annotation, {\textit{\remote}}, marks allocation sites to indicate that they have distinct lifetimes, sizes or variance among executions \todo{different from what}.
The first annotation,  \textit{\xsplit}, applied to procedure calls, indicates a call site that is likely to have different parallelism from the caller, \eg, the caller function executed in a single thread, but the callee executes in multiple parallel threads. Thus, the runtime should allocate different amounts of CPU resources for them. %\yiying{explain why at the call granularity (not smaller code scopes) and why at call sites not function def} 
We choose function call as the unit for annotating compute components as it captures parallel behaviors in most programs in practice, and this unit makes program analysis and transformation easier.
The second annotation, {\textit{\remote}}, marks allocation sites where objects have distinct lifetimes from other objects or with input-dependent sizes (\eg, an object's size grows with bigger program inputs), indicating potentially different memory resource allocation at runtime. 
Globally, users manage their spending using \texttt{@app\_limit(max\_cpu=K, max\_mem=N)} to limit peak CPU and memory for an application. 

{
\begin{figure}[t]
\begin{center}
\footnotesize
\begin{lstlisting}[
numbers=left,
xleftmargin=6.0ex,
frame=single,
framexleftmargin=15pt
]
# maximum number of vCPUs to use for the application
@app_limit(max_cpu = 10)

def group(df):
  return pd.groupby('COL1')
           .count(lamdba r: r['COL2'] == 'VALUE')
def sample(df): ...
  
def main(env):
  # a Dataframe whose size is input dependent
  @data dataset = pd.Dataframe()
  dataset.load_csv(env['INPUT'])
  counts, samples = [], []
  num_blocks = dataset.size // env['BLOCK_SIZE']
  
  for block in np.array_split(dataset, num_blocks):
    # num_blocks of calls execute in parallel
    counts.append(@compute group(block))
    samples.append(@compute sample(block))
    
  return samples, sum(counts)
  

    

\end{lstlisting}
% \vspace{-0.15in}
\mycaption{fig:code}{Example Python Porgram with \sys\ Annotation.}
{
This simple example loads a dataset, splits it into blocks, and performance counting and sampling on each block in parallel.
This annotation results in the resource graph shown in Figure~\ref{fig-materialize}. The \sys\ compiler transforms it into an internal representation shown in the \appendixautorefname.
}
\end{center}
% \vspace{-0.2in}
\end{figure}
}

Figure~\ref{fig:code} illustrates a simple code snippet that analyzes a dataset.
This example annotates the $pandas$ Dataframe object $dataset$ as a data component and the $group$ and $sample$ function calls as compute components. Both functions are called $num\_blocks$ times, and the annotation indicates that these calls can execute in parallel and the amount of parallelism is input-dependent. At the runtime, \sys\ decides the actual number of CPU cores to allocate, not to exceed 10 cores (specified at the top as $max\_cpu$), and the locations of the cores. For example, for one invocation, ten $group$ calls can run on five cores on the same server, while for another, four $group$ calls run on two cores on two separate servers.

Overall, these two annotations are easy to add, as developers usually know resource bottlenecks. To assist application porting, we further provide program profiling tools to help users identify potential locations to annotate.
Furthermore, \sys\ provides APIs for computation synchronization, explicit inter-component communication that developers can optionally use.
Overall, \sys's interface is simple, versatile, and generic, as it is not tied to any cloud providers.
% Overall, \sys's interface is simple and generic, and it can be built on top of different clouds~\cite{sky-computing}.

\subsection{Resource Graph and Its Generation}

After users submit their source programs to \sys\ and before their first and subsequent invocations, \sys\ prepares their execution under the hood by transforming user programs.
\sys's preparation centers around a new intermediate representation we propose called {\em resource graph}. 
Each graph node is a {\em compute} or {\em data} component, with the former being a code site with distinctive CPU usage patterns and the latter being a memory object with distinctive memory usage patterns. Each graph node has a resource feature that we record based on profiled history. Edges in a resource graph represent triggering or accessing relationships. 
%\zhiyuan{In addition to representing memory components sparately, \sys collects the resource lifetime and sizes in the resource graph.}
%\notezhiyuan{Explicit say what's different in a resource graph.}

\boldunderpara{Resource graph generation.}
Based on user annotations, \sys\ generates an initial resource graph where each \xsplit\ code site becomes a compute component, and each \remote\ object becomes a data component. \sys\ determines triggering or accessing edges by analyzing the program's control flow and data-accessing relationship.
%\zhiyuan{For parameters and return values only. We do not go into functions to find dependency.}
In addition to user-annotated data components, we perform program analysis similar to Mira~\cite{Mira} to identify the shared objects across compute components, which we separate as data components.

% As user annotations do not always reflect runtime resource usage patterns, \sys\ refines an initially generated resource graph with the help of profiling.
\sys\ samples an application's runs to capture the resource usage of each resource graph node (CPU usage for compute components, allocation size and lifetime for data components). It stores a histogram of all captured statistics with decaying weights at each resource graph node.
%The information will be used in proactive execution in \S\ref{sec:adaptive}.

% \sys\ also records the CPU amount (\ie, level of parallelism) and memory size for each refined component.
%\notezhiyuan{Should mention our identification is not should not be restricted to annotated componet.}
% After we acquire the resource graph for each profiling run, we unify all graphs by finding the finest splitting way across graphs.
% For example, suppose one profiling run results in the code split at lines 10 and 30, and another splits at line 20. We merge them by splitting at lines 10, 20, and 30. This combining method could lead to more components.
% Thanks to \sys's efficient component launching (\S\ref{sec:schedulemechanism}), locality-based scheduling (\S\ref{sec:schedulepolicy}), and scalable scheduler (\S\ref{sec:schedulemechanism}), it works well in practice. 

% As an optimization, \sys refines the graph and reducing the number of components by 
% grouping neighboring compute components into one component when they have similar CPU utilization, and pack multiple data components into one if their lifetime is similar in all history executions.

{
\begin{figure*}[th]
%\begin{minipage}{0.54\textwidth}
\begin{center}
\centerline{\includegraphics[width=\textwidth]{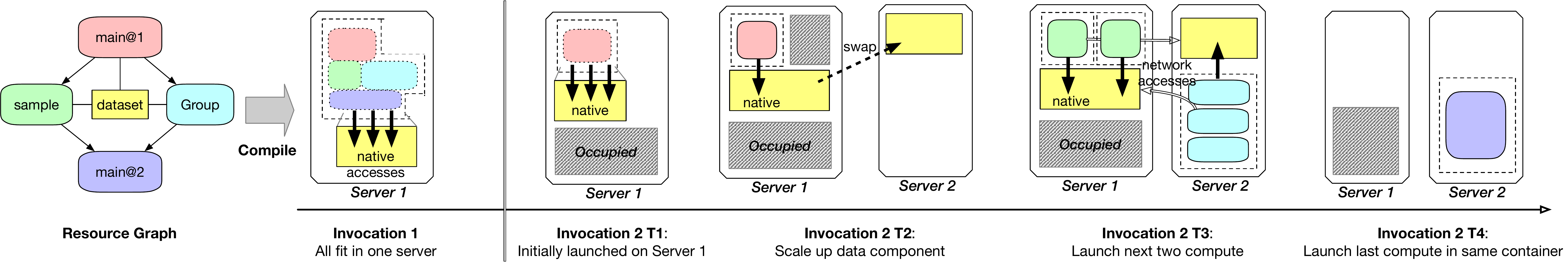}}
%\vspace{-0.1in}
\mycaption{fig-materialize}{Resource Graph and Adapative Execution.}
{
Corresponds to code example in Figure~\ref{fig:code}. Round boxes represent compute components. Rectangular boxes represent data components. Dashed boxes represent containers. Shadowed boxes represent occupied resources.
Invocation 1 runs everything on Server 1 which has enough resources at the time.
Invocation 2 runs on two servers as resources on Server 1 get occupied over time.
}
\end{center}
\vspace{-0.15in}
\end{figure*}

\boldunderpara{Compiling.}
\sys\ transforms a user program into internal forms for execution using the \sys\ compiler. 
Compilation happens primarily at the offline time, although the \sys\ runtime can also trigger the compilation of certain parts of the user program if needed.

Offline, the \sys\ compiler transfers each directed edge (one compute triggering another compute component) as an internal API that will be intercepted by \sys\ library at the runtime --- at this time, the \sys\ runtime can decide whether to continue the triggered component in the same execution environment or to launch new one(s).

Offline, \sys compiles two versions of each compute component that communicates with others (\eg, the top three computes in the graph in Figure~\ref{fig-materialize}).
The first version assumes that data components are local to all their accessing compute components, and we compile the accesses as native memory instructions. 
The second version assumes data components are placed on a different server from all their accessing compute components. 
For this version, we identify all memory accesses to data components and convert them into \sys's internal data-access APIs.
As frequent API calls into the \sys\ library incur high performance overhead, 
we optimize the generated code by batching accesses to multiple fields in a data structure as one API call to access a larger range.
%Mira explores more compiler optimizations for remote memory access; we leave exploring these optimizations for future work.
%Note that all interfaces have availability to runtime scale with a cost of performance. Details are described in \appendixautorefname.

As will be discussed in \S\ref{sec:schedulepolicy}, \sys\ dynamically determines the execution method for components in a resource graph. There can be cases where some compute components co-locate with some of their accessing data components, but remote to the others.
We don't try to precompile for all cases, as enumerating all combinations of local/remote accesses would result in explosive amounts of compilations.
Instead, we only pre-compile two versions (all local and all remote) and leave the compilation of the remaining cases to the runtime.
To improve performance, once the runtime compiles a version for one invocation, it is cached and reused for future invocations with the same component layouts.

\section{\sys\ Runtime System}
\label{sec:online}

This section discusses how the \sys\ runtime system schedules and executes an application's invocations with diverging resources requirements in an adaptive, proactive, scalable, and reliable way. 

\if 0
%\boldunderpara{Online system architecture.}
Figure~\ref{fig-arch} illustrates the overall architecture of \sys's runtime system.
Each rack has a {\em rack-level scheduler} for scheduling tasks within a rack, and the entire cluster has one global scheduler for assigning coarse-grained tasks to rack-level schedulers.
Each server has a {\em memory controller} for data component allocation and management, a {\em compute controller} for compute component execution,
%Controllers are responsible for creating, monitoring, and terminating their respective components. Additionally, each server has 
and a {\em network virtualization controller} for component communication orchestration. These three controllers sit in a user-space {\em executor} that interacts with the scheduler and other executors. 
For compute components, A {\em \sys\ runtime} is inserted within the application execution sandbox (Docker container in our implementation) to handle \sys\ internal APIs and monitoring tasks.
%for communication, manage a local memory buffer, and monitor resources and performance.
%This section discusses these components and how \sys\ adaptively materializes and executes applications. 
\fi
% \input{fig-runtime}

\subsection{Adaptive Scheduling and Execution}
\label{sec:adaptive}

To be adaptive, \sys\ innovates on scheduling policy and execution/auto-scaling mechanisms, with the overarching goal of achieving the best performance and resource efficiency for each invocation of bulky applications.

% \zhiyuan{better description for overall goal?}
%\sys aims to minimize the performance and resource overhead caused fine-grained component-level execution. \sys execute components that adapt to its need by \textit{materialize} the component in the graph: one component is physically executed as multiple physical components with runtime static size during execution.This materialized component is \textit{adaptive} to its execution.

%\sys auto-scales on demand by materializing to more physical components at runtime. This section describes the details in scheduling and executing components. 
%network communication and execution environment startup time while being resource-efficient.

\subsubsection{Adaptive Scheduling Policy}
\label{sec:schedulepolicy}

Overall, our scheduling policy tries to co-locate compute components and their accessed data components, as doing so avoids remote data accesses. It also tries to co-locate triggered and triggering compute components to avoid environment startup overhead and output/input messaging overhead.

Specifically, when scheduling an application's invocation, the scheduler first tries to find servers that can fit the entire application in the hope that all components can run in one environment. It chooses the server with the smallest available resources among them to leave more spacious servers for future larger invocations (\eg, invocation 1 in Figure~\ref{fig-materialize}). The scheduler marks the chosen server's with the potential need for the amount of resources of the whole application as estimated from profiled history. We do not allocate resources for future use in a resource graph. Instead, the marked resources are given low priority when the scheduler allocates resources for other applications. The chosen server starts the application in a container whose size is set according to the {\em first} compute component's profiled resource needs. If this compute component accesses data components, the server also allocates the data components locally if possible. Afterward, if this server still has enough resources for the next components, \sys\ chooses the same server and continues the computation in the same container process but changes the container size to what the new components need.

If there is not enough resources to co-locate components on any servers, \sys\ finds a server with the smallest available resources to launch each of them. For instance, in Figure~\ref{fig-materialize}, at invocation 2's T2, Server 1 runs out of resources, and \sys\ launches the blue component on Server 2. \sys\ then launches the remote-access compilation version or generates a new compilation on the fly if none exists.
%, which our library executes as network requests (\S\ref{sec:communication}).

When scaling up resources, \sys\ also follows the same locality policy. For example, \sys\ first tries to expand a data component on its current server. If that is not possible, \sys\ picks another server and prioritizes servers already running compute components that access the data component.  
%co-locate compute components and their accessing data components as much as possible to avoid network communication. We also try to schedule a subsequent compute component in the  the triggering and the triggered compute components as much as possible, as the former avoids network communication and the latter avoids environment startup time.

%To reduce the performance overhead of start-up containers and language runtimes for each function, there have been proposals to share these environments across functions.
Like \sys, some prior works also try to co-locate functions. \noteyiying{there should be some more papers on co-locating functions. fill more here}
A recent work, XFaaS~\cite{xFaaS}, proposes running multiple functions concurrently within one process and re-using runtime environments across invocations for serverless computing in the Meta private cloud. Differently, \sys\ does not co-run different applications, as our target environment is multi-tenant public clouds.
Moreover, unlike XFaaS, \sys\ is resource-centric instead of function-centric. 

\subsubsection{Adaptive Execution and Autoscaling}
\label{sec:execution}

We now discuss how \sys\ adaptively executes and scales data and compute components. 
Our high-level idea is to adapt the statically generated resource graph into physical execution units based on cluster availability and resource feature profiles. As such, multiple components in the original graph can be {\em materialized} as one {\em physical} component, while one original component can turn into multiple physical components.
The former can happen for two reasons: \sys groups neighboring compute/memory components into one when they have similar lifetime and scaling patterns over most history executions, and \sys\ co-locates components in one execution environment.
The latter happens when resource needs grow, and \sys\ automatically scales out a compute/memory component.

%\sys runtime focuses on providing close to native performance in all possible cases that accessing from remote during its lifetime. Our runtime ensures that through two key implementation: specialized, separated data component implementation and compiler assisted memory interface generation for compute component.

\boldunderpara{Compute component execution.}
\label{sec:computecomponent}
We execute compute components in containers (Docker in our current implementation) by initially sizing them according to profiling results and loading appropriate binaries (local or remote-access versions). %If the appropriate binary is not ready, we perform 
For the local-access version, compute components run directly on Docker. \sys\ only assists in mmap-ing local data components and forwarding the compute-component execution output to the scheduler. 
In the non-local mode, the \sys\ runtime executes remote-memory access APIs (which are generated by the \sys\ compiler) on top of an RDMA or TCP network stack.
%access APIs as local memory copies (for local data components) or network requests (for remote data components). 
%\sys\ waits for the reply from the rack-level scheduler before starting the subsequent compute components.
% \zhiyuan{Different interfaces for multiple possible remote accesses? \sys could remember history decision.}

\sys\ allows a compute component to use arbitrary amounts of memory or CPU as long as they do not exceed the user-specified limits. Thus, \sys\ may need to auto-scale compute components beyond the profiled configurations.
\sys\ detects when a compute component is under memory pressure and asks the scheduler to allocate additional memory space. The scheduler may schedule the additional space on the same or different server.
As such space is added after the launching of a compute component, \sys\ uses it as (local or remote) memory swap space and transparently performs swapping at the user level.
%\sys\ performs local memory swapping for adding remote physical component on local interfaces and local cache section for adding local component on remote interfaces. 

\boldunderpara{CPU resource autoscaling.} 
\sys\ supports both scaling up and scaling out.
For scaling up (and down), \sys\ simply changes a compute component's assigned vCPU. \sys\ periodically measures CPU utilization. If the utilization is 100\%, we increase vCPU by 1. %\fixme{add policy to reduce vCPU} %The final usage is captured, and the pattern is learned by \sys, as described in \S\ref{sec:policy}.
\sys's scaling out (and down) applies to compute components that have parallelism (such as $group$ and $sample$ in Figure~\ref{fig:code}).
\sys\ determines the vCPUs to use based on request load and use thresholds of adding/reducing vCPUs determined by history profiles.
For example, when an earlier invocation uses 10 vCPUs for 10 parallel execution of a compute component and has 50\% CPU utilization on all the vCPUs, a future invocation of 10 parallel execution would only use 5 vCPUs.
%\sys scales parallelism through mapping and merging from compute component invocations to parallel executions. As an example, in Figure~\ref{fig:code}, `transform` could be invoked 5 times at runtime, but \sys runtime decides to route the requests to three parallel containers.
%\sys use a request-rate based approach to decide the parallel unit, and the threshold is proactively decided by history execution.
%We follow the application-specified level of parallelism limits defined with $@app\_limit$ (\eg, $max\_cpu = N$ limits all parallel components to use up to $N$ vCPUs).
%Note that how to logically shard data is defined by the user program, as it is hard to automatically parallel execution without programmer effort.

%The \textbf{Appendix} presents the details of our \ufd-based swap system. 
%\zhiyuan{this two seems to be the same for me?}
%We implemented our swap system in user space using Linux's \ufd.
%It swaps to a remote space that internally maps to one or more physical memory components.
%These physical components are scheduled and executed in the same way as physical memory components for regular virtual memory components.
%The same swapping method applies to a mixed component if its run-time memory consumption exceeds the allocated amount.
%Our resource adjustment policy (\S\ref{sec-policy-component}) ensures components' initial value is correct and swapping would not affect the performance.

\boldunderpara{Data component launching and autoscaling.}
%\label{sec:memorycomponent}
\sys\ automatically starts a data component when the first compute component accessing it in the resource graph starts. %\noteyiying{when this compute component starts or when it first allocates memory?} 
The memory controller at the server chosen by the scheduler allocates a memory space with the size indicated in the resource graph. The data component ends when the last compute component accessing it ends or explicitly releases it. 

%\sys data components are always executed as isolated memory regions, regardless if it's locally or remotely accessed, and attach it to accessing compute components.
When a data component co-locates with a compute component accessing it on one server, the memory controller allocates a virtual memory space for the data component and $mmaps$ it to the compute-component container.
When a data component is not co-located, the memory controller manages it as a virtual memory space in a reserved \sys\ process for a TCP cluster or an RDMA memory region for an RDMA cluster, both of which are accessed by compute components via the network.
We explain how \sys\ isolates memory regions in \appendixautorefname{}.
%\noteyiying{how do we isolate different data components (local case and remote case)? need to talk about security and isolation here}

%\notezhiyuan{should mention physical here}
A data component can grow beyond its initially allocated size because of application's additional allocation.
%when application-level allocation happens.
In this case, the \sys\ memory controller on the server picked by the scheduler allocates an additional memory space.
%we dynamically add more memory spaces to automatically scale up the data component. Specifically, the \sys\ memory controller first tries to allocate more memory space on the same server where the data component currently lives. If there is no more space there, the memory controller asks the rack-level scheduler to find a proper server to allocate the new memory space. 
When a compute component is currently accessing a data component natively, \sys\ mmaps the additional memory space on the same node or performs swapping if the additional memory space is on a different node (T3 in Figure~\ref{fig-materialize}).
When a compute component is currently accessing the data component via \sys\ APIs, the \sys\ library will transparently issue network requests to multiple servers for accesses spanning separated memory spaces.

\subsection{Proactive Scheduling and Execution}
\label{sec:proactive}

Different from existing serverless systems that schedule and auto-scale resources on demand, \sys\ proactively schedules resources and sets up environments before resource change happens by leveraging profiled history, as discussed below.
%\zhiyuan{
%To reduce overhead of unprovisioned and fine-grained runtime scaling requests, \sys\ adopts a proactive approach. Leveraging historical profiling, \sys\ takes action before resource  change happens. \sys allocates and preserves resources ahead of time (\S\ref{sec:schedulepolicy}), pre-sets up execution and communication, and learns scaling patterns from history.
%}

%Scaling resources purely on-demand could cause two major performance problems: First, it could cause resource fragmentation. Means that the physical components for a single virtual component could be allocated on multiple machines due to the scaling at runtime. This happens especially when the resource amount varies across different executions. Second, frequent launching and scaling components could pressure scheduling and delay the execution. For example, the request-to-launch time in existing serverless system, even as low as 200ms \cite{??}, could be a huge overhead in \sys.

%We leverage the knowledge of history execution and action proactively to assist reactive auto-scaling at runtime. 

\subsubsection{Proactive Scheduling}

% \sys it sizes each component according to profiled history and adaptively scales it on demand.
% Together with our component execution mechanism (\S\ref{sec:execution}), \sys\ achieves both resource efficiency and performance efficiency.
Starting up the execution environment for a serverless application is a slow process that usually involves the start-up of a container/lightweight VM, language runtimes, and various libraries~\cite{shahrad2020serverless}. A single start-up time is not significant for long-running bulky applications, however, such an application can have a large resource graph and involve many rounds of start-up overhead. 
To avoid this overhead, we pre-launch a subsequent component in a resource graph when the current component is running. This pre-launching approach is also adopted by Orion~\cite{Orion-osdi22}, but Orion's pre-launching is based on user-supplied static serverless DAG, while \sys's resource graph is an intermediate representation that adapts dynamically.
Similar to traditional serverless frameworks~\cite{roy_asplos22,shahrad2020serverless,singhvi2021atoll}, we also pre-warm the first component in a resource graph based on historical invocation patterns.
%With the help of resource graphs, our scheduler can set the proper size of the subsequent component before its execution. This {\em just-in-time} component launching mechanism utilizes application resource behavior history, adapts to the most up-to-date rack resource availability, and hides scheduling and start-up overhead behind foreground performance. Note that a recent work, Orion~\cite{Orion-osdi22}, also pre-warms functions in a FaaS DAG but does not incorporate resource patterns.
% To be resource efficient, we target to make a scheduling decision for each node just before it is triggered so as to utilize the most up-to-date cluster resource availability information . 
%To further reduce start-up overhead, we pre-warm the first component in a resource graph based on historical invocation patterns, similar to previous works~\cite{roy_asplos22,shahrad2020serverless,singhvi2021atoll}.

{
\begin{figure}[t]
\begin{center}
\centerline{
\includegraphics[width=\columnwidth]{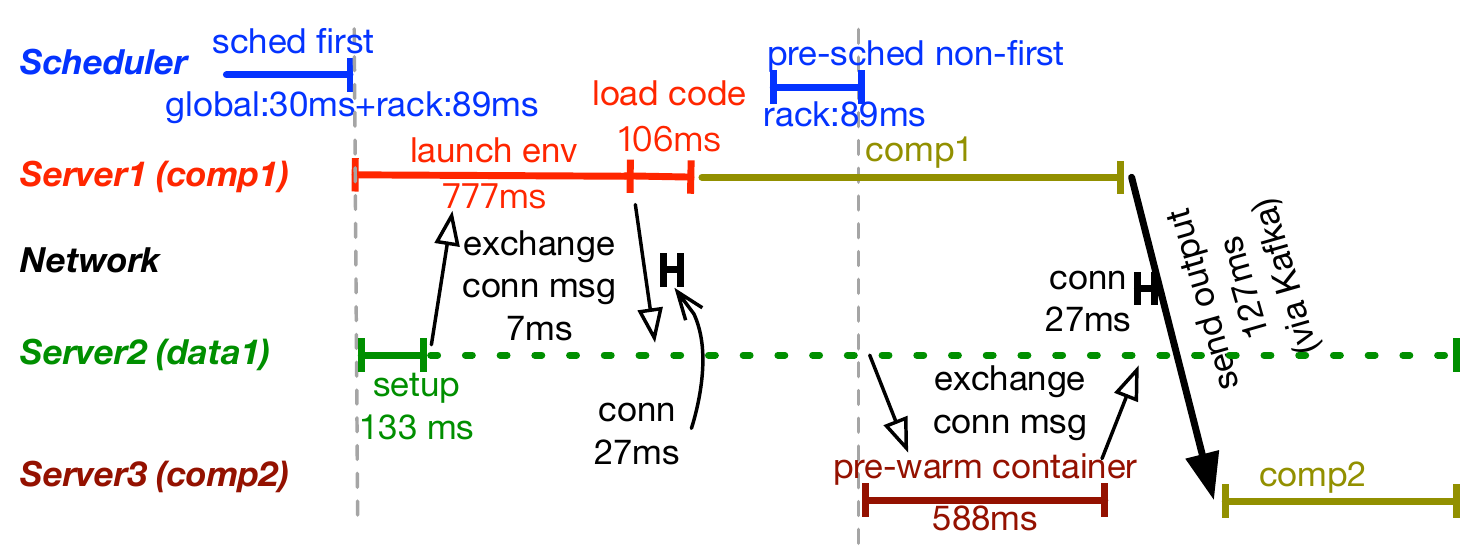}
}
\vspace{-0.1in}
\mycaption{fig-launch}{\sys\ Startup Flow with Real Results.}
{
This example application has two compute components accessing one data component; one compute triggers the other compute.
%Initially, \sys\ schedules the first compute and the data components.
%They then start their own launch processes in parallel.
%Once the environment is ready, they exchange their location information, after which their connection is established.
\sys\ pre-warms the environment for the second compute component and hides network connection set-up overhead behind performance-critical paths.
Dashed green line represents how long the data component lives.
%Dashed orange lines represent when network connection is up.
%\zhiyuan{Info exchange happens after components are inited, since the exchanged message is the QP info not the address info. And its async. So the process can be memory sent message to proxy after memory init, and computes sent its info after compute init, this two arrows happens async. May need to change the graph.}
}
\end{center}
%\vspace{-0.2in}
\end{figure}

\

\subsubsection{Asynchronous Communication Setup}

\label{sec:communication}

As communicating or accessing components in a resource graph can sit on different servers, especially in a highly utilized rack, it is crucial to ensure that both the actual communication (data plane) and the connection setup (control plane) are fast.
We support both RDMA and TCP communication, with several optimizations on the control path and the data path. %Below, we explain these optimizations, with the Appendix present our detailed RDMA implementation. Most of the optimizations except for zero copy apply to TCP as well.
Similar to existing remote-memory works~\cite{Mira}, we optimize the data path by batching small network requests and by caching fetched data locally. For RDMA, we perform one-sided, zero-copy communication\noteyiying{true?}.

The communication control-path optimization is unique to serverless computing, as we need to minimize network connection set up time, while allowing the direct communication between components in the data plane. 
The general challenge in serverless communication is that a function cannot easily know where other functions are.
Prior serverless communication systems rely on an overlay network to build connections~\cite{thomas2020particle}) or pre-establishes connections between all pairs of endpoints~\cite{copik2021rfaas,wei2022booting,wei2021krcore,mohan2019agile}. The former is slow, while the latter does not meet our resource elasticity goal. Both try to establish connections via {\em direct} channels between two nodes.

Our idea is to leverage {\em non-direct but already established} connections to assist the initial location discovery.
Our observation is that a component has always established a connection with the scheduler by the time it starts up.
Moreover, the scheduler is the one who decides and thus knows the physical locations of the two components that will be communicating. % (\ie, communicating relationship in the component graph).
We let the scheduler send one component's physical location (in the form of executor ID) to its communicating component's {\em network virtualization controller} when initiating the two components.
With the location information, the two components can easily establish a datapath connection (\eg, RDMA QP).
To further optimize the startup performance, we initiate the location exchange process as soon as the execution environment is ready while the user code is loaded in parallel. Doing so hides the location-exchange overhead behind the performance-critical path (Figure~\ref{fig-launch}).

\subsubsection{History-Based Resource Adjustment}
\label{sec:policy}

To improve the overall performance of executing applications in a \sys\ cluster, we dynamically set the initial allocation size and the adjustment amount.
Rather than adjusting components based solely on current metrics, we incorporate historical usage to avoid over-adjusting for one-time input changes.

\sys decides how to scale each component with two parameters, \textit{initial size} and \textit{incremental size}. \sys\ tunes these values to strike a balance between performance and resource efficiency.
The \textit{initial size} of a component decides the resource to allocate at the start-up time.
During runtime, \sys autoscales the resource amount one \textit{incremental size} at a time to avoid frequent small resource adjustments.
\sys\ re-adjusts these two sizes periodically after $K$ (\eg, 1000) executions of an application.
For compute components, we have initial and incremental sizes for both vCPU amount and memory amount. For memory components, we have sizes for memory amount.
We describe the detailed tuning method in \appendixautorefname{}.

%For compute components, we adjust two parameters for a number of allocated vCPUs.
%For both compute and memory components, we adjust their memory allocation sizes. We detail the algorithm of deciding the values in \appendixautorefname{}.

\subsection{Scalablity, Reliability, and Consistency}
\label{sec:reliability}
We end the design section by discussing how \sys\ scales its scheduler, handles failure, and supports consistency.

\subsubsection{Scalable two-level scheduler}
As a bulky application often has multiple scheduling entities (components in a resource graph), as well as frequent runtime scale to adapt to inputs. \sys\ needs a scalable scheduling mechanism. We propose a two-level scheduler architecture where a cluster has one {\em global scheduler}, and each rack within has a {\em rack-level scheduler}.
When a user event triggers an application, the request is sent to the global scheduler (possibly after a data-center load balancer directs requests across multiple clusters). The global scheduler maintains the rough amount of available resources in each rack. It uses this information to direct the application request to a rack by balancing loads across racks. It then looks up the application's compilations in the compilation database and sends the compilation and corresponding resource graph to the rack-level scheduler. 

The rack-level scheduler handles allocation requests for each component in a resource graph. It chooses a server to initially run the application based on policies in \S\ref{sec:schedulepolicy}. Upon completing a compute component, its result is sent to the rack-level scheduler via reliable messages (\S\ref{sec:failure}). The release of a data component or the need to scale it up are also sent to the rack-level scheduler. The rack-level scheduler increases the server's available resources accordingly and decides where to schedule the subsequent component or the scaled-up part of a component. It then updates the resource counting of the destination server and notifies that server's executor to start the component. Thus, the rack-level scheduler always has an accurate view of available resources in all the servers in the rack and what components currently run on each of them.
%maintains the available resources per server and what components currently run on each server. 
If a rack runs out of resources for a request, the rack-level scheduler sends the request back to the global scheduler to find another available rack.

{
\begin{figure*}[th]
\begin{minipage}{0.48\columnwidth}
\begin{center}
\centerline{\includegraphics[width=\columnwidth]{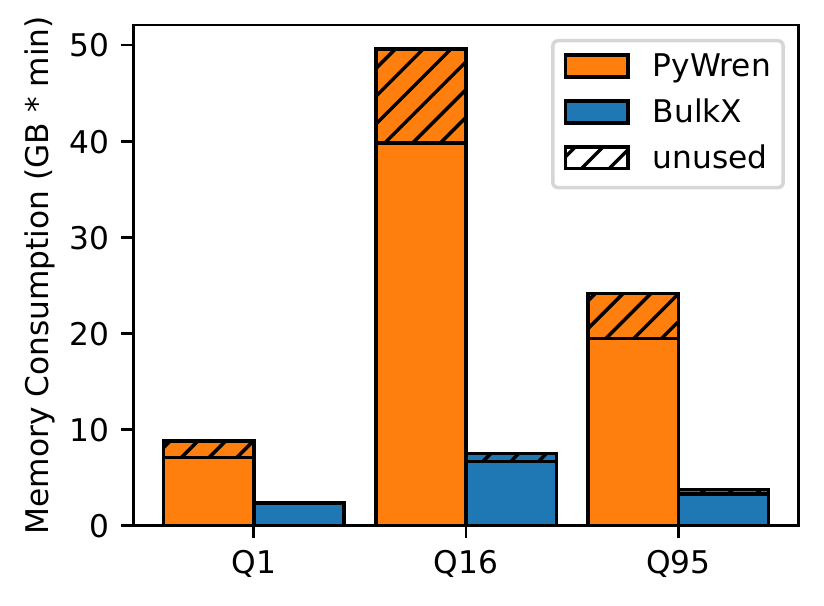}}
\vspace{-0.1in}
\mycaption{fig-tpcds-resource}{TPC-DS Memory Consumption.}{Running Query 1, 16, 95.}
\end{center}
\end{minipage}
\hspace{0.015\columnwidth}
\begin{minipage}{0.48\columnwidth}
\begin{center}
\centerline{\includegraphics[width=\columnwidth]{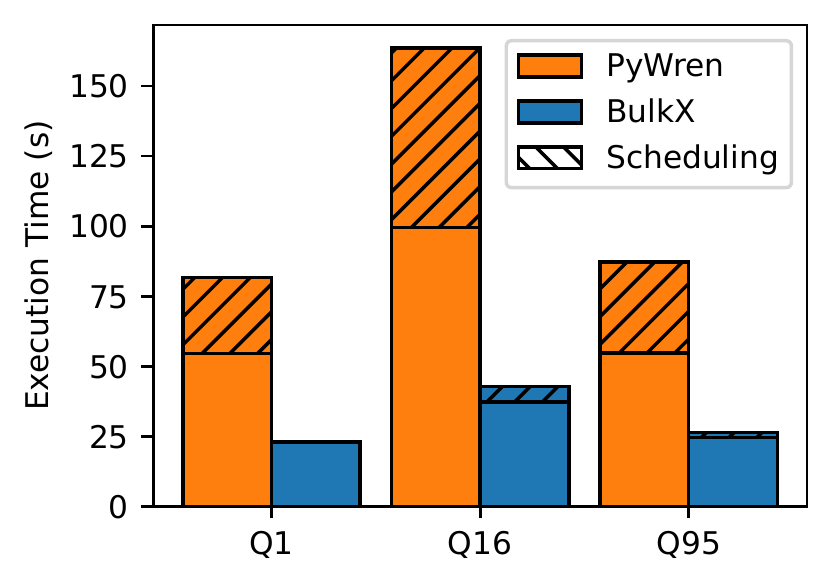}}
\vspace{-0.1in}
\mycaption{fig-tpcds-time}{TPC-SD Execution Time.}
{
Running Query 1, 16, 95.
}
\end{center}
\end{minipage}
\hspace{0.015\columnwidth}
\begin{minipage}{0.48\columnwidth}
\begin{center}
\centerline{\includegraphics[width=\columnwidth]{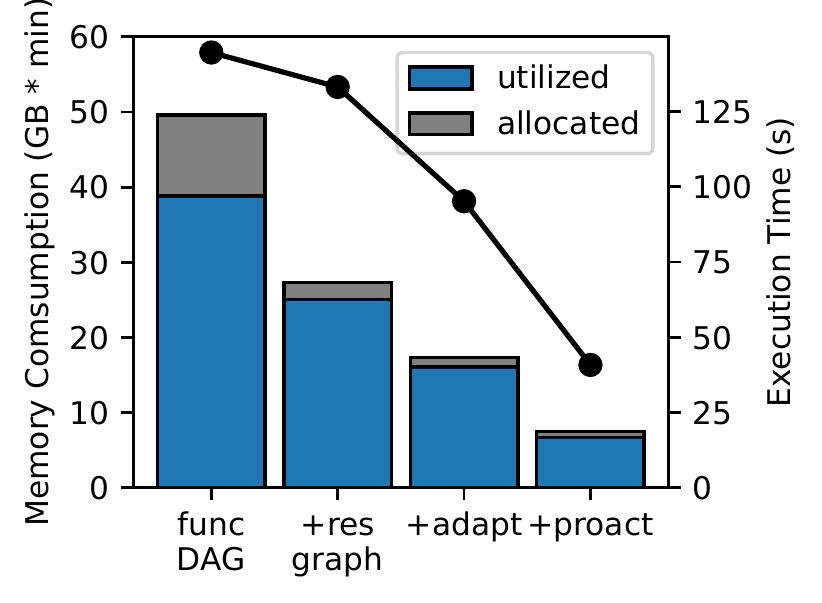}}
\vspace{-0.08in}
\mycaption{fig:eval:breakdown-tpc}{Resource and Performance Breakdown for TPC-DS Query 16.}
{
}
\end{center}
\end{minipage}
\hspace{0.015\columnwidth}
\begin{minipage}{0.48\columnwidth}
\begin{center}
\centerline{\includegraphics[width=\columnwidth]{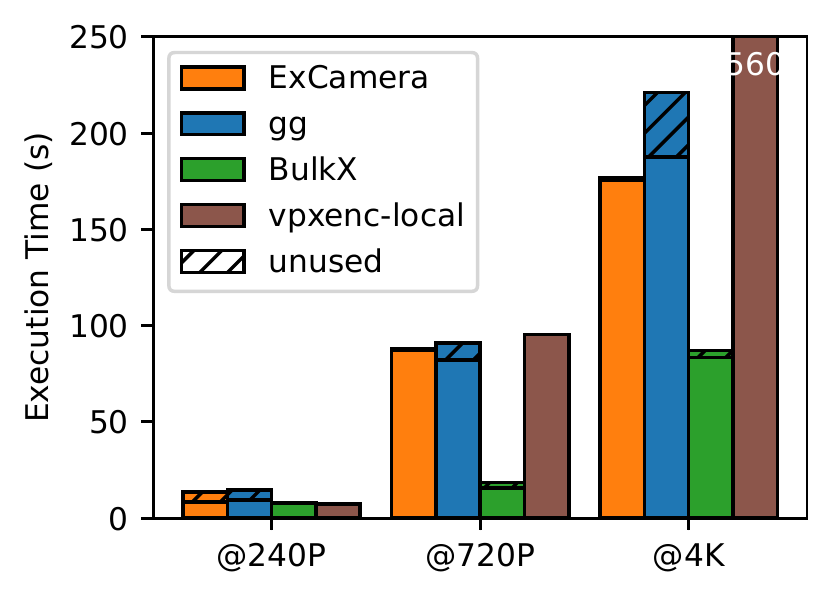}}
\vspace{-0.05in}
\mycaption{fig-excamera-time}{Video Transcoding Execution Time.}
{
With 3 video resolutions.
}
\end{center}
\end{minipage}
\end{figure*}
}

\subsubsection{Failure Handling}
\label{sec:failure}
Traditional FaaS-based serverless re-executes the entire function when failure happens. This works well when applications are small, but bulky applications can be time- and resource-consuming to re-execute. To avoid re-executing an entire bulky application after a failure, we send the result of every compute component to the rack-level scheduler via reliable messaging (\eg, Kafka~\cite{Kafka} in our implementation).
When a failure happens, \sys\ discards the crashed component and all data components it accesses.
\sys\ locates the latest cut of a resource graph where all the edges crossing the cut have been persistently recorded.
%, all other physical components belonging to the same virtual component, and all this virtual component's communicating components. 
\sys\ re-executes from this cut using the persistently recorded input messages. 

We apply the same failure-handling principle to data components. 
\sys\ does not make data components durable or replicate them, as they only store intermediate state~\cite{klimovic2018pocket}. If any memory region in a data component crashes, \sys\ discards all the compute components that access it and all other memory regions of the same data component. \sys\ then finds the latest reliable resource graph cut to restart.

Resuming from the reliably recorded messages in the above manner is sufficient to guarantee {\em at-least-once} semantics, which is the same as today's serverless systems~\cite{AF_reliable_event_processing_20,GCF_exec_guarantees_2022}. 
%If no restarted component modifies the external state, we can further guarantee {\em observably exactly-once} semantics.
Our failure handling design balances foreground performance and failure recovery performance compared to mechanisms that take more checkpoints~\cite{Burckhardt_2021} or those that restart the entire function chain after a failure~\cite{AWS-Step-Functions,OpenWhisk-composer}.

\subsubsection{Consistency and Synchronization}
Multiple compute components can access the same data component as shared memory. \sys\ provides basic distributed synchronization primitives instead of a particular consistency scheme. Users can use the \sys\ synchronization primitives like distributed locks to enforce critical sections.
As \sys\ runs on generic RDMA or TCP network, all communication is through messaging, and we do not provide automatic coherence across components. New coherent media like CXL could potentially be used with \sys\ to provide hardware-level coherence, which we leave for future work.

\section{Evaluation}
\label{sec:results}

\boldunderpara{Implementation.}
We implemented \sys's compiler and program analysis on top of Mira~\cite{Mira}.
We implemented \sys's runtime system on top of Apache OpenWhisk~\cite{OpenWhisk}.
In total, \sys\ includes $\mathtt{\sim}$19.6K SLOC: 3.2K lines of change on Mira, 10K lines of Scala to extend Openwhisk's scheduler and executor, and 4.6K lines of C together with 1.8K lines of Python for runtime and libraries.
Currently, \sys supports programs written in Python, the second most commonly used programming language in serverless~\cite{Datadog_2023}, and in C++, a common language used for bulky applications.

\boldunderpara{Environment.}
We evaluated \sys\ on a local rack of a Dell PowerEdge R740 servers each equipped with two 16 core Intel Xeon Gold 5128 CPU, 64\GB\ of memory, and a 100\gbps\ Mellanox Connect-X5 NIC. 
We use eight servers for running \sys\ and systems in comparison and four additional servers to run Redis.
All the servers are connected to a 100\gbps\ FS N8560-32C 32-port switch. All servers run Ubuntu v20.04 with Linux v5.4.0. We use Docker v24.0.7 for containers.

\subsection{End-to-End Application Results}
\label{sec:eval-app}

We first present the high-level results of running three bulky applications on \sys: data analytics with Pandas~\cite{pandas} running the TPC-DS benchmark~\cite{tpcds:sigmod02}, video processing pipelines, and simple machine-learning training using logistic regression.

{
\begin{figure*}[th]
\begin{minipage}{\figWidthSix}
\begin{center}
\centerline{\includegraphics[width=\columnwidth]{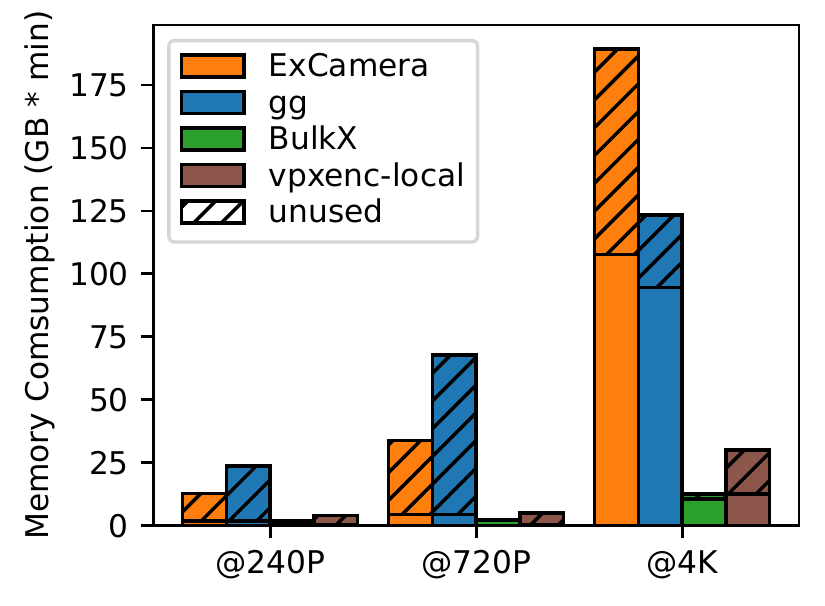}}
\vspace{-0.1in}
\mycaption{fig-excamera-mem}{Video Transcoding Memory Consumption.}
{
}
\end{center}
\end{minipage}
\begin{minipage}{\figWidthSix}
\begin{center}
\centerline{\includegraphics[width=\columnwidth]{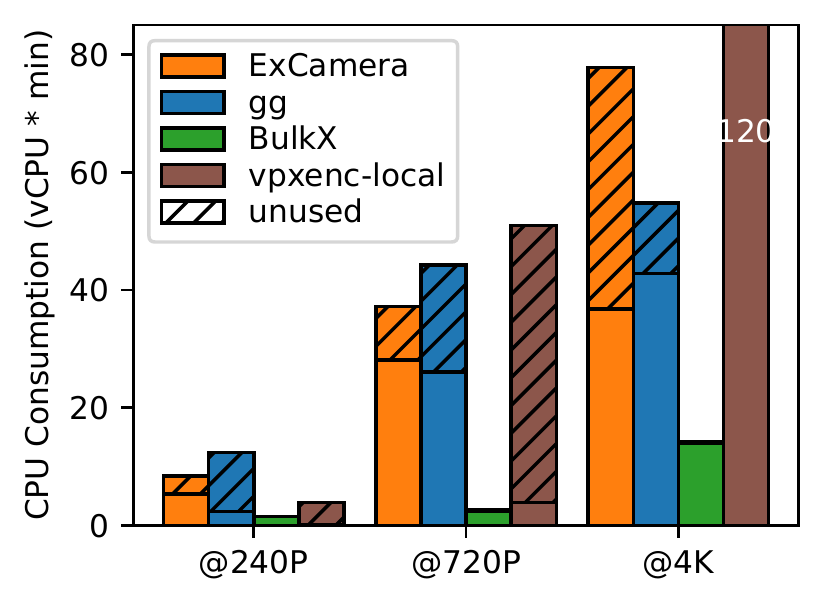}}
\vspace{-0.1in}
\mycaption{fig-excamera-cpu}{Video Transcoding CPU Consumption.}
{
}
\end{center}
\end{minipage}
\hspace{0.005\columnwidth}
\begin{minipage}{0.48\columnwidth}
\begin{center}
\centerline{\includegraphics[width=\columnwidth]{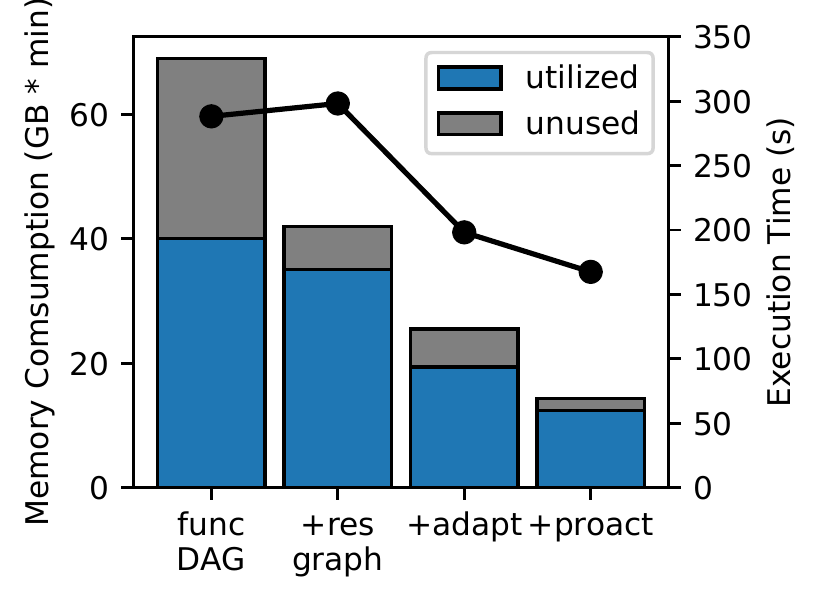}}
\vspace{-0.1in}
\mycaption{fig:eval:breakdown-excamera}{Resource and Performance Breakdown for 4K Video Transcoding.}
{
}
\end{center}
\end{minipage}
\hspace{0.005\columnwidth}
\begin{minipage}{\figWidthSix}
\begin{center}
\centerline{\includegraphics[width=\columnwidth]{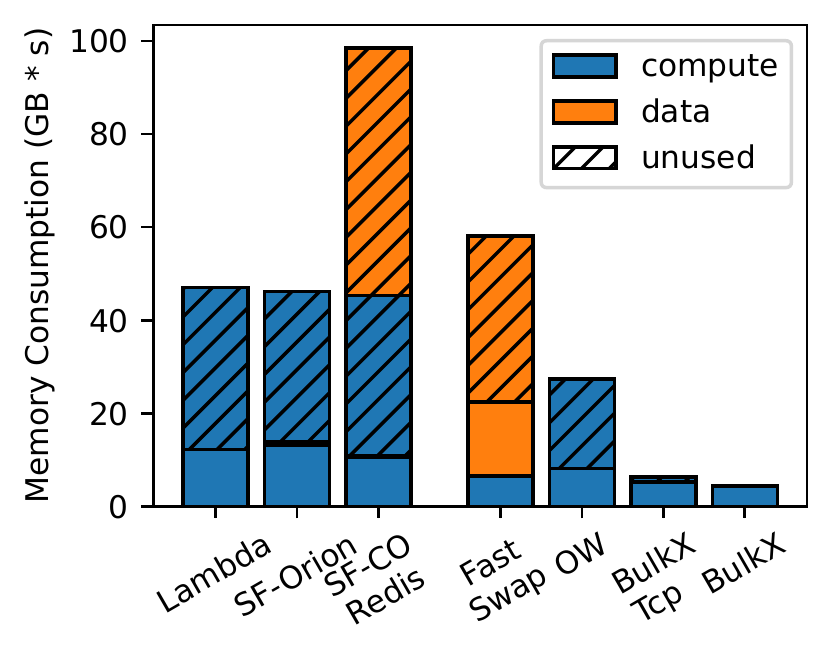}}
\vspace{-0.1in}
\mycaption{fig-lr-resource-mid}{LR Memory Comsumption with Small Input.}{}
\end{center}
\end{minipage}
\end{figure*}
}

%\centerline{\includegraphics[width=\columnwidth]{Figures/Figure-TPCDS-independent-computescale.pdf}}
%\vspace{-0.1in}
%\mycaption{fig-tpcds-computescale}{Independent Compute Element Scaling in TPCDS.}

% \zhiyuan{Two key points. Explains why resource is more efficient. Explain the performance gain (mainly adaptive + cluster level). Performance reason is different for different inputs?}

\subsubsection{Distributed Query with TPC-DS}
\label{sec:eval-tpc}

% \zhiyuan{Describing the requests}

The first application we evaluate is distributed data analytics on Pandas running the TPC-DS benchmark~\cite{tpcds:sigmod02}.
We compare \sys\ with PyWren~\cite{jonas2017occupy_pywren} running on OpenWhisk on our local cluster.
%6-9 stages already
PyWren splits each TPC-DS query into several stages, each containing some DataFrame operators. For each stage, PyWren generates different number of workers to perform the computation, and we use Orion~\cite{Orion-osdi22} to set the resource for each worker. Intermediate data storage for PyWren uses Redis servers in the cluster. 
%\zhiyuan{We use Orion to tune resource for each stage.}
We implement each query as a single Python program and annotations.
% \sys infers data components when converting each query to a resource graph.

%To port this application, we use th8  e \sys\ DataFrame compiler and library (\S\ref{sec:porting}) to convert DataFrame operators into compute components and accessed DataFrame objects into data components. Since with our library, DataFrame operators can directly access remote memory as computation progresses, we remove the explicit data fetching/storing steps in Pywren (except for the first and the last operator which still loads/stores data from Redis hosting the original database).

We use multiple input dataset, whose size ranging from 2\GB\ to 1\TB\, to evaluate three TPC-DS workloads with different performance characteristics and varying resource requirements: queries 1, 16, and 95.
They read 2.5\GB, 20\GB, and 19\GB\ of data respectively and have peak memory usage of 240\GB\ and peak CPU of 120 vCPUs.

\boldunderpara{Overall cost and performance.}
Figure~\ref{fig-tpcds-resource} shows \sys's and PyWren's total memory consumption. PyWren failed to scale efficiently with multiple isolated functions. \sys\ reduces memory consumption by 72.5\% to 84.8\%.
PyWren provisions each stage for its peak memory usage.
Moreover, each worker in PyWren fetches all the data it will access from Redis before the computation starts and stores the data back after all its computation finishes. Essentially, PyWren pays for the same amount of memory consumption twice.
In contrast, \sys\ right-sizes each component in a single stage by auto-scaling on demand. Moreover, \sys\ does not incur the double-memory-consumption problem,
Both systems achieve max CPU utilization of 120 cores. In average, \sys utilize 91.2\% of CPU time, much higher compared to PyWren's 63.8\%, as function-centric scale causes resource stranding.
Figure~\ref{fig-tpcds-time} plots the total run time of the three queries.
\sys\ outperforms PyWren by 54.2\% to 63.5\%, primarily due to adaptive native execution and faster parallel compute instance execution. With adaptation, over 84\%, 78\%, and 82\% components for Q1, Q16, and Q95 co-locate on the same servers.
% Moreover, PyWren incurs significant I/O performance overhead because of its serialization/deserialization. In comparison, \sys's accesses to a remote data component is via zero-copy RDMA with no need for serialization. 
Comparing across all different queries, Q16 has the largest resource consumption reduction and performance improvement because it has a higher degree of parallelism and a more complex sharing pattern, which \sys\ handles a lot better than PyWren.

\boldunderpara{Ablation study.}
To understand the effect of different \sys techniques, we add one technique at a time and show the resulting memory consumption and application performance in Figure~\ref{fig:eval:breakdown-tpc} using the TPCDS-16 query.
We start with the baseline of a static function DAG as used by PyWren.
We then use our static resource graph instead of the original function DAG and execute each graph node in a separate environment. Thanks to a resource-oriented decomposition, by scaling on resources instead of functions, it greatly reduces resource consumption both in used and unused memory.
% \yiying{is our resource graph more coarse grained than function DAG?} \zhiyuan{no}
It slightly improves application performance, but still has relative high remote access and scheduling overhead.
%First, going from function-centric to resource centric allows resource to be shared between multiple computation components and scale resources independently. This greatly improves the resource efficiency, especially reduces the allocated but not used resource caused by resource bundling. However, it also introduces extra scheduling overhead and the remote access could be inefficient remote access. In TPCDS queries, since sharing memory also reduces the serialization overhead, resource-graph based execution has better performance.
Afterward, we add the adaptive scheduling and execution support (\S\ref{sec:adaptive}), which co-locates many graph components and use local communication interfaces, thus greatly improves application performance.
Lastly, we add the support of proactive scheduling/execution and history-based resource adjustment (\S\ref{sec:proactive}). The former improves performance because of environment/connection pre-setup. The latter improves resource efficiency because of more accurate resource sizing and less runtime remote scales.

{
\begin{figure*}[th]
\begin{minipage}{0.48\columnwidth}
\begin{center}
\centerline{\includegraphics[width=\columnwidth]{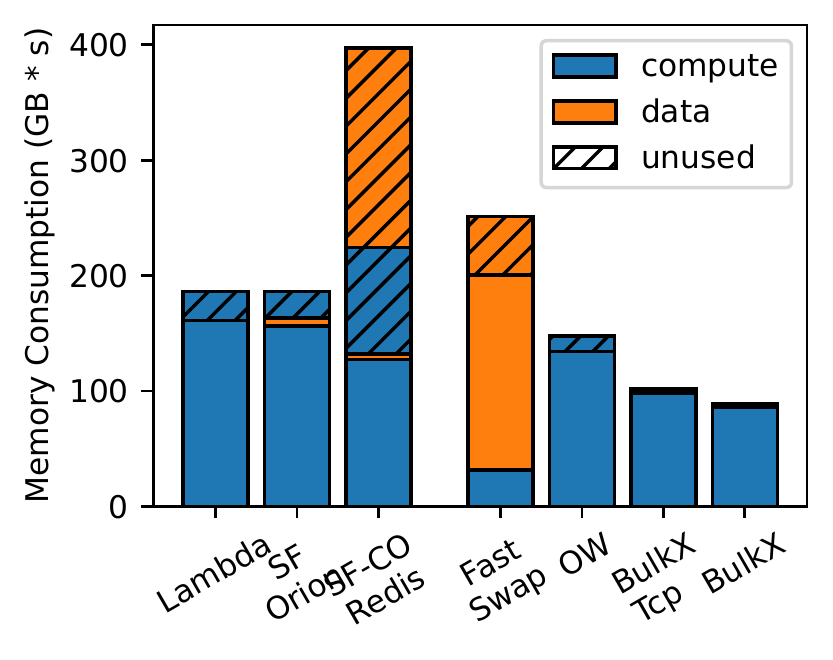}}
\vspace{-0.1in}
\mycaption{fig-lr-resource-large}{LR Memory Comsumption with Large Input.}{}
\end{center}
\end{minipage}
\hspace{0.005\columnwidth}
\begin{minipage}{0.48\columnwidth}
\begin{center}
\centerline{\includegraphics[width=\columnwidth]{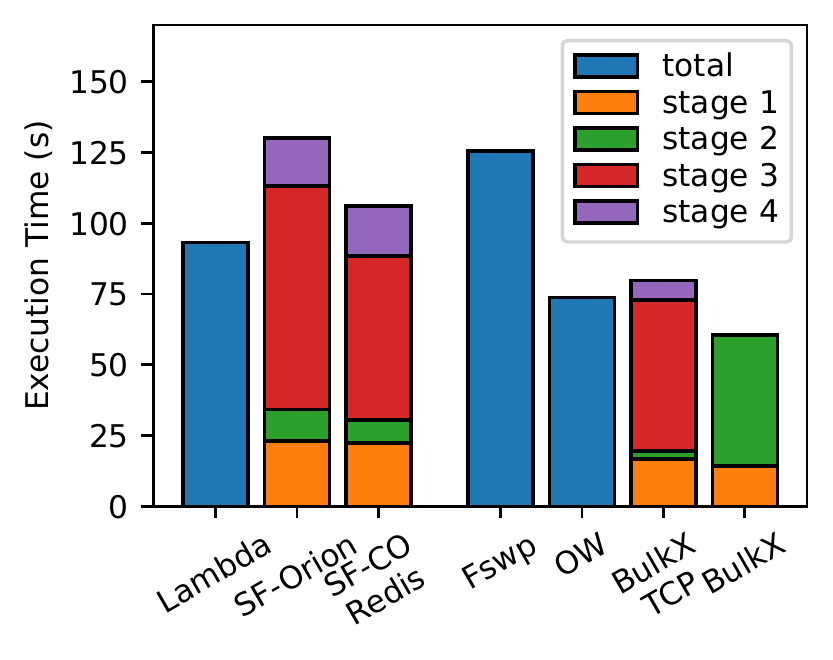}}
\vspace{-0.1in}
\mycaption{fig-lr-time}{LR Execution Time Breakdown with Large Input.}{}
\end{center}
\end{minipage}
\begin{minipage}{0.48\columnwidth}
\begin{center}
\centerline{\includegraphics[width=\columnwidth]{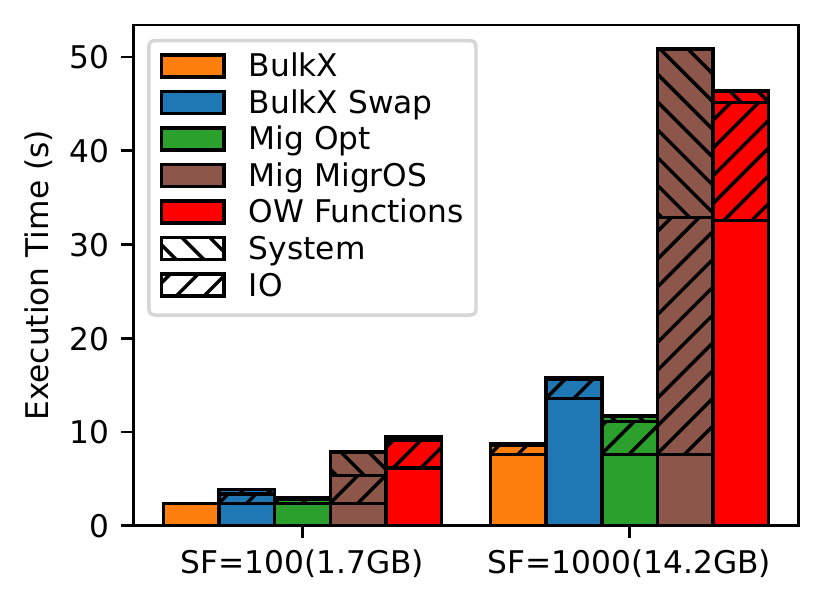}}
\vspace{-0.1in}
\mycaption{fig-tpcds-flex}{Comparison of Runtime Scaling Technologies.}
{
System: schedule \& migrate preparation.
}
\end{center}
\end{minipage}
\hspace{0.02\columnwidth}
\begin{minipage}{\figWidthSix}
\begin{center}
\centerline{\includegraphics[width=\columnwidth]{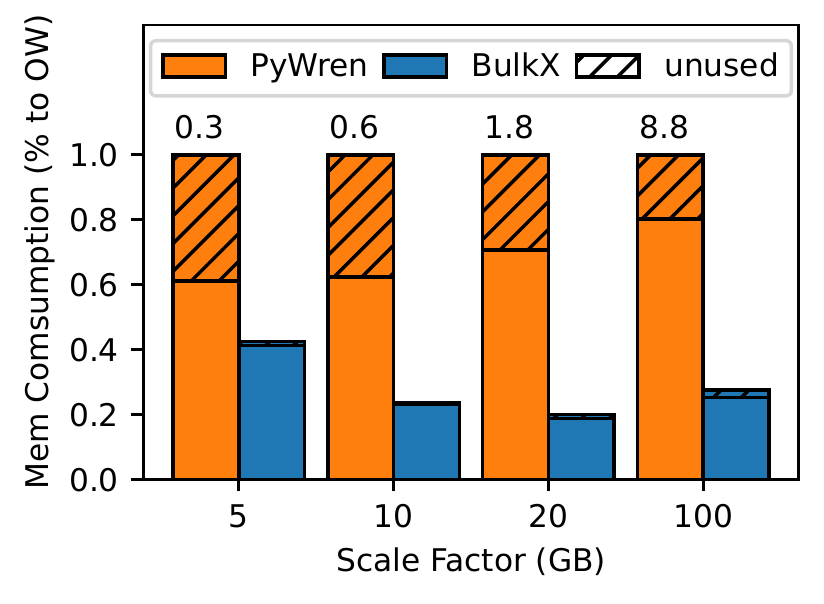}}
\vspace{-0.1in}
\mycaption{fig-tpcds-input}{Memory Consumption of TPC-DS Q1 against Input Sizes.}
{
Bars for relative values, numbers absolute.
}
\end{center}
\end{minipage}
\end{figure*}
}

\subsubsection{Video Processing}
\label{sec:video}

Our second application is video processing, where we perform video transcoding using the encoding/decoding operators developed by ExCamera~\cite{excamera}. ExCamera divides video frames into batches and uses one parallel worker to encode a batch.
Our version is a single program with 11 annotations, where input video are sliced into parallel segments and each segment are processed with up to 16 parallel compute units. We use ExCamera's set up of six frames as a single encoding unit and a batch size of 16 units. 
\sys\ generates a resource graph with 37 compute and 33 data components.

We compare \sys\ to ExCamera and ExCamera running on the gg~\cite{fouladi2019laptop} serverless frameworks. Original ExCamera uses a fixed VM to schedule and merge results, while running decoding steps parallel in serverless workers, while gg is a pure serverless implementation. The gg implementation represents the encoding of each frame batch with 80 functions, each reading from and writing to a Redis intermediate storage pool. 
We also compare with a local $vpxenc$ encoder~\cite{vpxenc} that runs everything natively on one server in our cluster.

\boldunderpara{Overall cost and performance.}
Figure~\ref{fig-excamera-time} plots the total execution time of the three systems processing a 1-minute slice of the movie ``Sintel''~\cite{sintel2010} in three resolutions: 240P, 720P, and 4K. The peak resource usage hits the cluster resource limit of 120 CPUs and 174GB of peak memory in both Openwhisk and \sys. \sys\ achieves the best performance for all three resolutions. \sys\ performs adaptive materialization and launches 81.3\% of components in same batch on the same server. In contrast, even when servers used for the gg experiments have the same amount of free resources, gg's functions share data that always lives on separate Redis servers.
$vpxenc$ runs in one server, but it is limited by the amount of parallelism it can achieve: only 18 out of 32 allocated cores and 14 out of 16GB allocated memory are actually used.
This difference is more apparent with larger videos that benefit from larger parallelism.

%However gg fails to achieve resource- and performance-efficiency as it scales up.
Figure~\ref{fig-excamera-mem} and Figure~\ref{fig-excamera-cpu} plot the total memory and CPU consumption. 
\sys\ consumes the least amount of memory.
As function-based frameworks, ExCamera and gg set the same function size for all inputs. Thus, it has to make provisions for the largest anticipated inputs, causing significant unused resources for the smaller 240P and 720P videos.
% For the largest 4K video, gg incurs significant execution-time overhead and memory overhead for reading/writing data to Redis, which putting together, causes high memory consumption. 
$vpxenc$ has similar or lower used memory as \sys, but its unused memory amount is much larger. As a single-unit execution, its size needs to be set to the peak memory size and cannot be adapted to requirement change over time.

Similar to memory, ExCamera, gg, and $vpxenc$ also incurs high CPU resource wastage because of their peak-based static provisioning. Unlike memory, their used CPU resource is also much higher than \sys\ with 720P and 4K resolutions, mainly because of their longer execution time. %\noteyiying{I don't get why longer execution time doesn't result in larger used memory for them, esp. gg 720P}

\boldunderpara{Ablation study.}
We perform a similar ablation study on video processing, as shown in Figure~\ref{fig:eval:breakdown-excamera}. Similar to TPC-DS, each added technique reduces memory consumption for video processing. Differently, video processing incurs high overhead in scaling memory objects with small increments, which causes performance downgrade when simply changing from function DAG to resource graph. Performance is greatly improved when adaptive and proactive optimizations are added.

%GG uses a single type of worker to execute multiple types of requests, so in serverless system it has to do peak provision (allocate resource for largest possible work) and resource waste, as shown in Figure~\ref{}.
%, which incurs large overhead in the execution on both scheduling (esp. with smaller inputs) and execution part (esp. with larger inputs).

%Also, knowing the data passing relationship enables adaptive materialization, which reduces all remote access overhead.

%In real execution, over 83\% of almost all batches are over 98\% of them finished the execution on one machine. Our adaptive materialization converts the data sharing to shared memory access in those cases. All components that cannot be accommodated on the local machine found slot on the sibling machines. This approach could scale to thousands of workers. 

%\boldunderpara{Providing resource flexibility efficiently.}
%We provide both memory and CPU flexiblity for bulky vpxenc. scale to 120 concurrent cores and 174GBs of peak memory usage.

%we are performant under different cases:
%We have comparable performance and resource utilization with local version on smaller data. serverless function incurs scheduling and execution overhead.

%We reduce the data sharing overhead on larger cases.

% \input{fig-cluster-level}

\subsubsection{Logistic Regression}
\label{sec:eval-lr}

THe third application is machine learning training using logistic regression (LR) ported from Cirrus~\cite{carreira2019cirrus}. Here, we focus on comparing with a variety of serverless baselines.
It first loads the input data set and separates it into a training set and a validation set.
Then, it performs the regression on the training set.
Finally, it validates the trained model with a validation set.
%
%We annotate the original code with 3 \xsplit\ annotation points where tasks change. 
\sys's resource graph for LR includes four compute components corresponding to the above steps and three data components (training set, validation set, and learned weights).

We compare \sys\ (RDMA and TCP) with executing the original program on vanilla OpenWhisk and on FastSwap~\cite{FastSwap}, a system optimized for remote-memory swapping. OpenWhisk executes the application as one function sized to the peak memory size. FastSwap uses the same amount of local memory as \sys's compute component and remote memory of the peak memory size. We run these systems on our local cluster to have controlled network environments. 
We also ran a set of configurations on AWS: 1) running the entire program as one Lambda function, 2) running the same four code pieces as \sys\ compute components in four Lambdas orchestrated by AWS Step Functions~\cite{AWS-Step-Functions} and storing data components in S3 or Redis~\cite{REDIS}. %running on an EC2 t3.medium instance. 
%As Redis is faster than all existing serverless computing frameworks with intermediate storage pool~\cite{klimovic2018pocket,sreekanti2020cloudburst}, we only evaluate Redis here.
For Step Functions,  we use two settings: %\textit{{SF-PO} (performance optimized)}: size Lambda functions to achieve the optimal performance; 
\textit{{SF-CO} (cost optimized)}: size Lambda to achieve the lowest cost~\cite{power-tuning,Eismann_2021,COSE}; % with Redis used for intermediate state; %(in GB$\times$second, which is proportional to AWS's billing of Lambda but does not use dollar price); 
and \textit{{SF-Orion}}: size each function using Orion~\cite{Orion-osdi22}.
%Although we only compare with AWS, the problems we show prevail in other clouds.

Figures~\ref{fig-lr-resource-mid} and \ref{fig-lr-resource-large} plot the total memory consumption of these schemes when processing two input dataset sizes, 12\MB\ and 44\MB, which result in peak memory usage of 0.78\GB\ and 2.4\GB\ respectively. 
%We separate consumption accounting between the compute side (Lambda, \sys\ compute components, FastSwap's local node) and the memory side (S3, Redis, \sys\ data components, FastSwap's remote memory).
%\notezhiyuan{need to explain colors. May be "resource wastage" first? }
As all schemes use the same number of vCPUs in this experiment, CPU consumption comparison directly translates to comparing execution time, which we show in Figure~\ref{fig-lr-time} for the 44\MB\ input.
Overall, \sys\ consistently achieves the lowest resource consumption, highest resource utilization, and best performance, and the improvement is higher with the small input. 
%For the two inputs, \sys\ (when running on RDMA) reduces resource consumption by 40\% and 84\% compared to OpenWhisk while only adding 1.3\% performance overhead. 
Even when running on TCP, \sys\ still largely reduces resource consumption with small performance overhead.
%Between the two inputs, the smaller wastes more resources for the baseline systems but not \sys.

%Moreover, \sys\ only has 0.9\%-3\% unused resources, while OpenWhisk wastes 9\%-70\%.
Among local baselines, FastSwap has the highest resource consumption and performance overhead among the three, because it wastes a significant amount of remote memory by allocating for the peak (no autoscaling) and its swap-based coarse-grained remote memory access overhead. 
Comparing the AWS baselines, running the same resource graph on Step Functions only reduces 2\%-5\% memory resource consumption with SF-CO and SF-Orion compared to single Lambda, far from how much \sys\ reduces over OpenWhisk.
Digging deeper, we found that splitting a program into Lambda functions introduces both performance and resource overhead, as seen in Figure~\ref{fig-lr-time}.
Each function needs extra time to read/write data from S3 or Redis and serialize/deserialize the data. Serialization and deserialization also requires extra memory space. 
%Moreover, the communication between Lambda and S3 is slow.
%However, communication speed to S3 is not the main performance overhead, as can be seen from the only-slightly-better run time when running on Redis, an in-memory key-value store. 
These results demonstrate that simply writing an application as function DAGs on today's serverless platforms is not enough. % for achieving cost and performance goals.
%\notezhiyuan{splitting into components cannot achieve performance and resource efficient at the same time}
%\boldunderpara{Resource wastage on AWS.}
%Figures~\ref{fig-lr-resource-mid} and \ref{fig-lr-resource-large} also report allocated but unused resources (\ie, resource wastage). 
Additionally, all AWS baselines have huge resource wastage, especially with the small input, for several reasons. %There are multiple causes of resource wastage on AWS.
FaaS services only allow one function size for all invocations and throughout an invocation's execution.
Moreover, AWS fixes the CPU-to-memory size ratio, causing waste of CPU or memory.
%memory is wasted when a function reads/writes data from storage and serialize/deserialize them.
%Fourth, CPU and memory sizes have fixed ratio. When optimizing for performance, the CPU size fits the needs but memory will be over-provisioned, causing more resource wastage in SF-PO. In contrast, when we set the size to be minimal memory, execution time increases significantly (not shown in the figures).
Finally, the long-running Redis instance is provisioned for peak resource usage and wastes huge amounts of resources at non-peak times.

\subsection{Closer Looks}
\label{sec:eval-micro}

To further understand \sys, we performed a set of additional evaluations to focus on various features of \sys.

{
\begin{figure*}[th]
\hspace{0.01\columnwidth}
\begin{minipage}{0.49\columnwidth}
\begin{center}
\centerline{\includegraphics[width=\columnwidth]{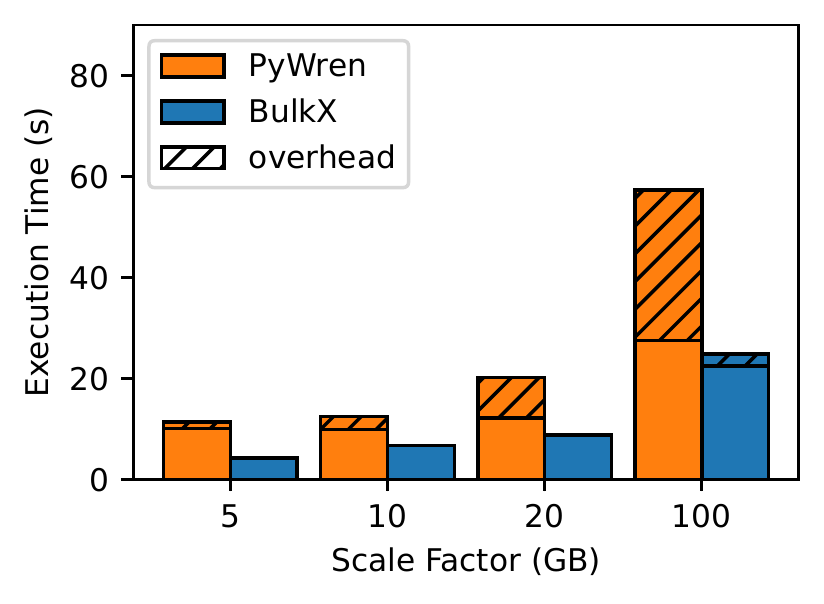}}
\vspace{-0.1in}
\mycaption{fig-tpcds-input-time}{Execution Time of TPC-DS Q1 against Input Sizes.}
{ \\
}
\end{center}
\end{minipage}
\hspace{0.015\columnwidth}
\begin{minipage}{0.49\columnwidth}
\begin{center}
\centerline{\includegraphics[width=\columnwidth]{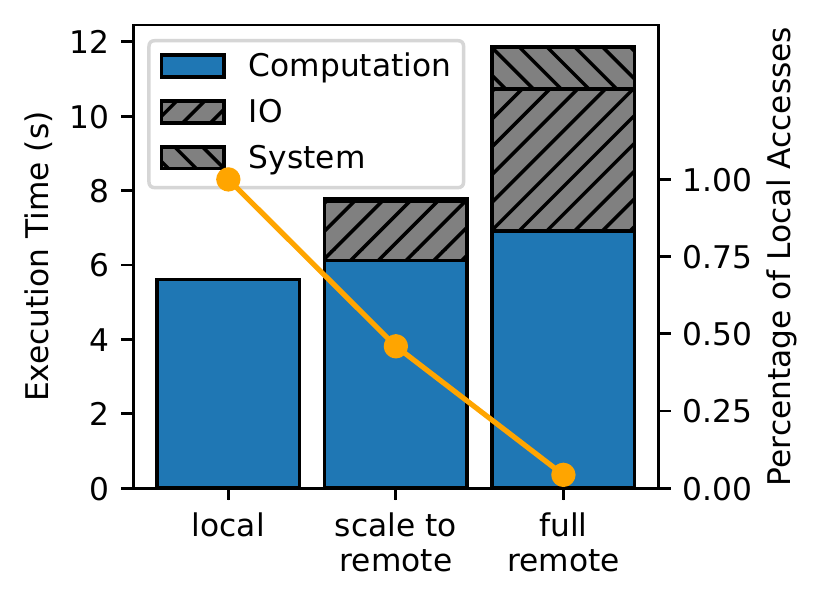}}
\vspace{-0.1in}
\mycaption{fig-tpcds-adaptive}{Effect of Locality-Based Placements.}
{ \\
}
\end{center}
\end{minipage}
\hspace{0.015\columnwidth}
\begin{minipage}{0.48\columnwidth}
\begin{center}
\centerline{\includegraphics[width=\columnwidth]{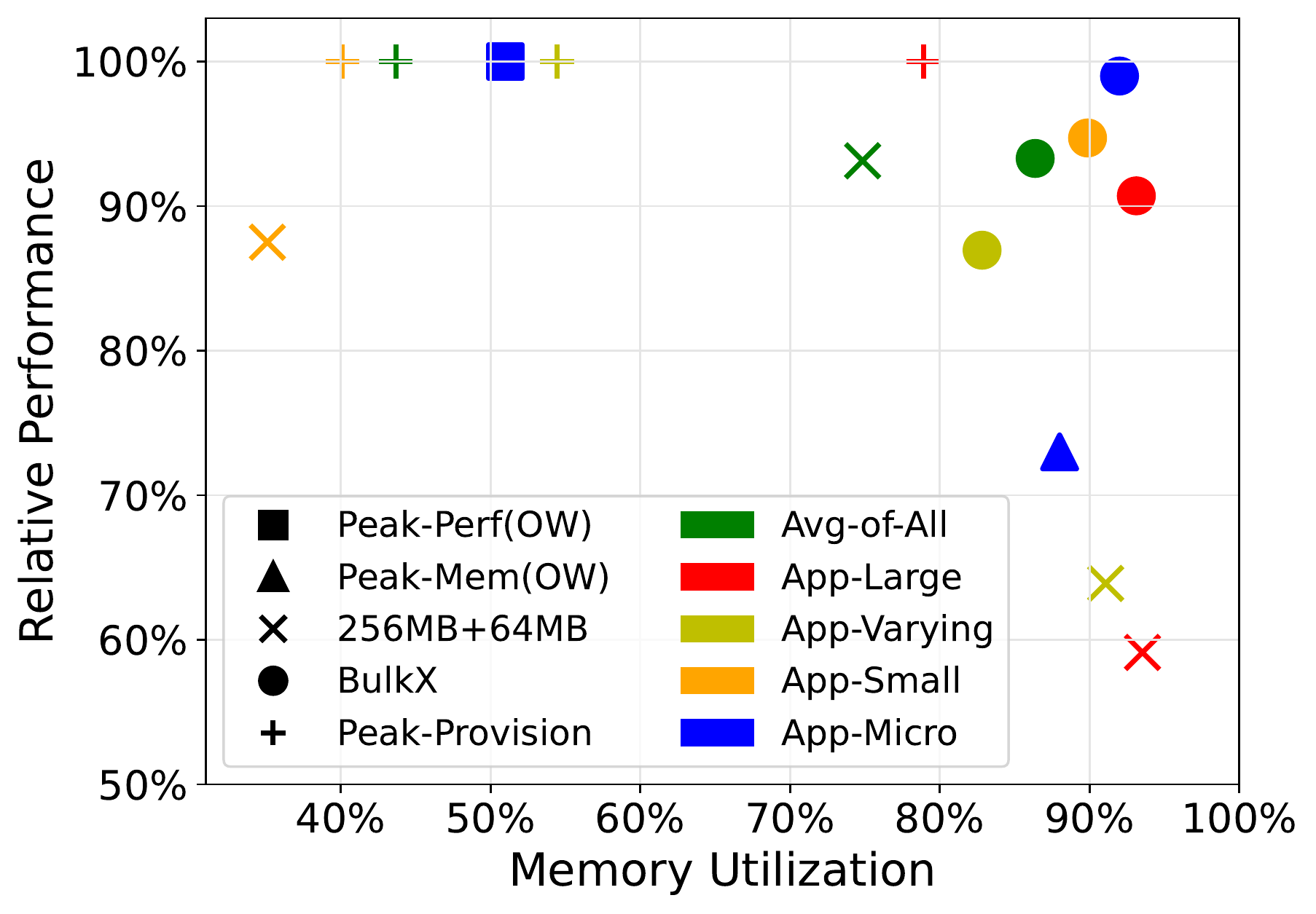}}
\vspace{-0.1in}
\mycaption{fig-policy-adjustment}{Performance and Resource Utilization with Different Sizing Strategies.}
{
}
\end{center}
\end{minipage}
\hspace{0.015\columnwidth}
\begin{minipage}{0.48\columnwidth}
\begin{center}
\centerline{\includegraphics[width=\columnwidth]{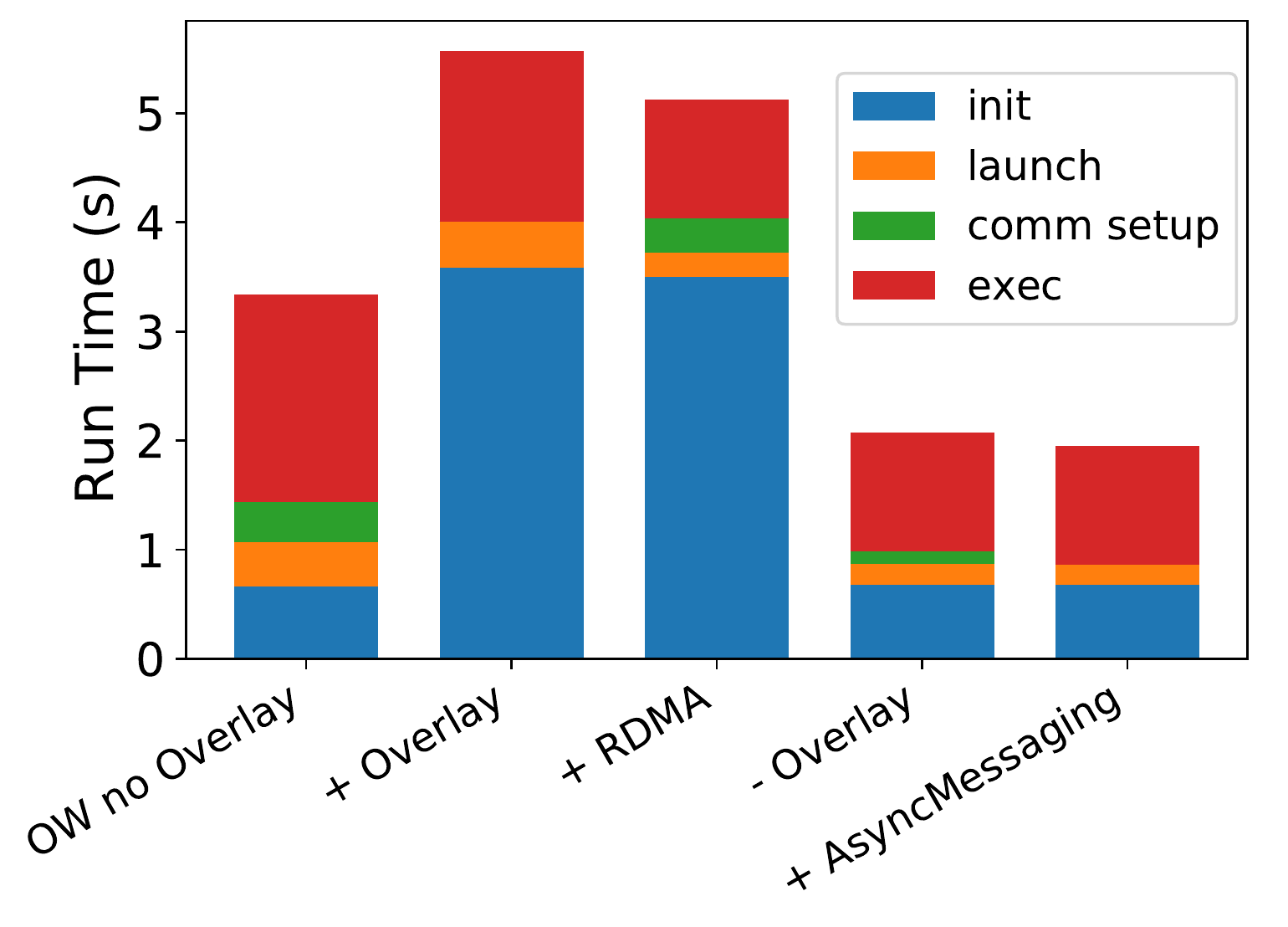}}
\vspace{-0.1in}
\mycaption{fig-launch}{Effect of \sys Communication Startup Techonologies}
{}
\end{center}
\end{minipage}
\end{figure*}
}

%\centerline{\includegraphics[width=\columnwidth]{Figures/Figure-TPCDS-independent-computescale.pdf}}
%\vspace{-0.1in}
%\mycaption{fig-tpcds-computescale}{Independent Compute Element Scaling in TPCDS.}

\boldunderpara{Alternative runtime scaling methods.}
As discussed in \S\ref{sec:disagg}, resource disaggregation and migration are two possible ways of adjusting resources and avoiding wastage.
We compare \sys\ with 1) swap-based disaggregation (to be fair, we run all \sys's systems except that we perform swapping for all accesses beyond local memory), 2) a best case for live migration by only counting pure data movement time using full 100\gbps\ network bandwidth, 3) migration with MigrOS~\cite{migros}, and 4) vanilla OpenWhisk.
For a fair comparison, we let all systems use the same amount of total resources (not counting unused resources).

As shown in Figure~\ref{fig-tpcds-flex}, for different inputs, the size of $Join$ stage varies from 267\MB\ to 14.7\GB.
with \sys's adaptive materialization, all components run locally with the scaling factor 100, and 34\% of data components' memory regions are scheduled at remote servers with the scaling factor 1000.
Disaggregation solutions like the swap results we show always place data in remote servers, causing high application-execution performance overhead because of network communication.
With migration-based solutions, the application execution runs natively and has no overhead. However, migrations cause significant overhead even with an optimal best case, because bulky applications' memory consumption is high (SF 1000).
Finally, OpenWhisk incurs significant performance overhead for the reasons covered before.

%To compare with disaggregation, we 
% \zhiyuan{In this exmaple we focus on the scaling of the `JOIN` operator in TPCDS-16 quries, which joins three different tables. Depends on the input size, the required memory from a single worker in this operator could go from 176MB to 14.6GB.}
%fig-eval\_tpcds\_same\_resource\_flexibility: Compare with different resource-scaling policy. we use the same allowed resource amount, and same runtime scaling policy (memory), means that no over-allocated is allowed. \sys executes the memory as a.
%Note that migration cannot scale resoruce lively, it will migrate from one fixed size to another. In this case we simulate migrate 
%\zhiyuan{More comments on the migration: migration is not designed for scale multiple times. The figure show one time scale up cost, but in real application the scale up can happen more than one time, along with the program execution. Under this senario our system still pays similar overhead, but a migration-based appeaoch need to migrate multiple times.}

\boldunderpara{Adaptation to different inputs.}
To understand how \sys\ adapts to different inputs, we use the TPC-DS Query 1 with five input sizes (5, 10, 20, 100, and 200\GB) to compare \sys\ with PyWren. 
Figure~\ref{fig-tpcds-input} shows the memory consumption comparison in relative terms, with absolute numbers reported on the top (CPU consumption is similar and not included).
\sys's memory consumption is consistently lower than PyWren. \sys\ adapts its component sizes at the run time according to inputs, thus resulting in negligible resource wastage. PyWren uses one function size across inputs, resulting in significant resource wastage, especially when inputs are small. Its used memory is also higher than \sys\ because of the reasons discussed in \S\ref{sec:eval-tpc}.

Figure~\ref{fig-tpcds-input-time} plots the corresponding execution time. \sys\ achieves good performance, especially with small input sizes, thanks to its adaptive-materialization. \sys\ is able to schedule everything locally for scale factors 5 and 10, 96\% local for SF 20, and 95\% of parallel compute components distributed evenly across four servers. \sys\ is able to pack most non-shared data components local to the compute components, with the shared data components accessed remotely.
% During the execution, most memory accesses goes into one scalable data component (for example, in stage 2, randomlly accessed, scales from 62M to 1.32G) that are placed in the local node, and only data compoentns sharing between computes (scaling from 12MB to 46MB) are accessed remotely. 

%eval\_tpcds\_scale\_resource and eval\_tpcds\_scale\_data shows how \sys adapts to different inputs. on TPCDS-1 Query. With resource-scaling, it keeps a high resource usage for all input size; With adaptive materialization, it keeps high performance for all cases. It could be divided into 3 phases, full local, partial local and divided.

%On performance side, with adaptive materialization, Mira \cite{} identifies the access to memory elements and generate the best case memory access.

\boldunderpara{Adaptive placement.}
To compare different adaptive placements of \sys, we run the $ReduceBy$ operator within TPC-DS Q16. It involves a set (3 to 120) of parallel compute components passing their result to a single compute component (fan-in), with one data component used for data sharing for each sending compute component (total data size from 730\MB\ to 113\GB). 
Figure~\ref{fig-tpcds-adaptive} shows three cases: local runs everything in one machine, remote-scale scales up data components to a remote machine (with 46\% being local), and disagg runs all components on different machines. Application execution time increases with more components being remote, and most of the overhead comes from pure I/O movement. The system overhead of setting up execution environments and network connections is small.

%Each each worker have one randomly accessed data comp and another data comp for sharing data with reducer. 
%Dependes on the input, the number of parallel workers ranging from 3 to 120 and size of memory component ranging from 730MB to 113GB.
%eval\_tpcds\_adaptive\_materialization. local: All-in-one machine.
%remote access : The randomally accessed data comp is scaled onto the other machine.
%disaggregation of components. almost all physical elements are disaggregated.

\boldunderpara{Proactive resource adjustment.}
Figure~\ref{fig-policy-adjustment} plots the performance and memory utilization with three schemes: fixing initial and incremental sizes to 256\MB\ and 64\MB, provision data components to the highest usage (peak-provision), and \sys's history-based size setting. 
We use real-world serverless application memory usage profiles from Azure~\cite{shahrad2020serverless}.
%This dataset contains a set of real-world applications, each being invoked many times with different user inputs.
In addition to the average of all the datasets, we choose a few representative applications to highlight.
% Figure~\ref{fig-micro-adjust-data} plots the CDF of memory usage across invocations of these representative applications and the average memory usage of all applications in the trace.
%The representative applications are 
\textit{Small}: most invocations' memory usages are small, \textit{Large}: most invocations used a lot of memory, and \textit{Varying}: different invocations' memory usages vary largely.
Detailed invocation resource usage distributions could be found in \appendixautorefname{}.
Overall, \sys\ achieves high utilization and good performance.
Peak provisioning achieves the best performance since no auto-scaling or swapping is needed, but has low memory utilization. 
%\zhiyuan{A fixed configuration only works well when an application's usage pattern happens to be close to the configuration. actually this cannot process variance data. Maybe just remove this} 
A fixed size configuration can lead to both poor resource utilization (when real usage is below the configuration) 
and poor performance. (when frequently adding many physical memory components). % at runtime).

\boldunderpara{Scheduler scalability.}
Our evaluation shows that \sys's global scheduler can process 50k application invocations per second, 2.5 times faster than OpenWhisk's cluster scheduler. The rack-level scheduler can schedule 20k components per second. We further analyze the component processing rate a rack can execute. Our evaluated applications have an average of around 1 second duration per component; each (compute) component consumes one core on average. A rack typically has less than 1,000 cores. This means a rack’s computation rate is 1,000 components per second. Applying the same type of calculation using today's serverless function duration and resource consumption~\cite{shahrad2020serverless}, we get 9,804 functions per second to be a rack's computation rate.
%If we use today’s serverless functions to model components, according to~\cite{shahrad2020serverless}, each function’s average duration is around 0.6 seconds; each function consumes avg 128MB memory which translates to 0.17 cores using AWS Lambda’s ratio. This means a rack’s processing rate is 9804. 
Both computation rates are much smaller than our scheduler’s 20,000/s request rate. This analysis is robust regardless of the number of components in an application. %, as the more components there are in an application, the more processing needed (fewer applications but same amount of components to schedule).
Overall, \sys\ can sustain a cluster of around 100 to 1000 racks, which is more scalable than OpenWhisk. As enable resource sharing between clusters already achieves high resource utilization, future implementations could deploy multiple global level scheduler to achieve better scalability and performance. 
% To scale further, one could change our current single-threaded implementation of the global and rack schedulers to be multithreaded and/or deploy multiple global schedulers.

\boldunderpara{Proactive execution and network setup.}
Figure~\ref{fig-launch} shows the impact of \sys's communication techniques on a workload with one compute component accessing one data component in warm started environments with no established connections. The first bar represents OpenWhisk without an overlay network, while the second bar adds the overlay network to enable direct component-to-component communication. The third bar replaces the TCP stack in the second bar with \sys's RDMA stack, improving component execution time. The fourth bar removes the overlay network and uses \sys's network virtualization module, reducing initialization time. Finally, the fifth bar includes \sys's asynchronous messaging optimization, hiding network connection setup time.

\section{Conclusion}
%\vspace{-0.05in}
We presented \sys, a resource-centric serverless framework that executes bulky applications adaptively. Our thorough evaluation results demonstrate \sys's performance and resource efficiency over existing function-based serverless frameworks. 
We hope \sys\ can allow more types of applications to port to and benefit from serverless computing.
Future works can extend \sys\ by adding more features like more versatile consistency support, more comprehensive compiler features, and more programming language support. 
%on a disaggregated environment.
%based on the ideas of resource disaggregation and aggregation.
%We show that disaggregation can be used to lift the limits of current serverless platforms while still maintaining comparable or even better performance. 
%Our evaluation of \sys\ demonstrates its benefits in resource saving and performance improvement. 
%\fi

%\clearpage
\bibliographystyle{plain}
\bibliography{references}

\clearpage
\pagenumbering{arabic}
\twocolumn[
\begin{@twocolumnfalse}
\begin{center}
{\Large\bf Appendix}
\end{center}
\smallskip

\bigskip
\end{@twocolumnfalse}
]

{
\begin{figure}[th]
\begin{center}
\footnotesize
\begin{lstlisting}[
numbers=left,
xleftmargin=6.0ex,
frame=single,
framexleftmargin=15pt
]
# The compiled "group" component with remote
# memory component access
def _remote_group(context):
  df = context.get_remote_addr('df')
  # Mira optimized remote access
  _cacheline = _prefetch(df.buffer(), N)
  for _el in _el_per_line:
    _v = _cacheline[_el:_el+1]
    ...
  # release the remote usages.
  df.release()
  return sum

# simplified invocation API implementation
def invoke(function, rt, *args, **kwargs):
    # read cached proactive scheduling
    # result or request scheduling
    exec = rt.schedule(func)
    # if decide to execute inside a container
    if exec.location == LOCAL:
        return function(*args, kwargs)
    # if decide to execute remotely
    elif exec.location == REMOTE:
        # need to pack and pass required enviroments
        exec.set_params(...)
        exec.pack_environment(...)
        rt.launch(exec)
        

\end{lstlisting}
% \vspace{-0.15in}
\mycaption{fig-api}{\sys\ Compiled Python Code.}
{
Describes how \sys intercept function call APIs and how it generate remote accesses.
}
\end{center}
% \vspace{-0.2in}
\end{figure}
}

\section{Additional Programming Model Details}

We now provide additional details of our programming model.

\subsection{Programming Model and APIs}

\sys supports most heap objects in c++ and certain types of memory objects as memory components in python.
For c++ program, we allow all global variables and heap objects allocated through new operator and $malloc$ function call to be memory objects. For python program, we support python standard data structures including list and dict, as well as objects follows buffer protocol and array protocol. Most popular data objects in python, including numpy $array$, pandas $dataframe$ and pytorch $tensor$ implemented the protocols and wide range of applications on machine learning, data analytic, batched data processing are supported by \sys.

In addition to annotations, \sys provide APIs for inter component communication and synchronization. All @compute calls are required to have a return value. In compute components could call `@message` to a return value of a `@compute` call to send the message, call `@mutex` to acquire and hold a lock and call `@barrier` to synchronize the execution. Internally, \sys use the messaging controller and proxy network to efficient implement the primitives.

\subsection{Adaptive Compilation and Generated Memory Interface}

When compiling into resource graph, \sys identifies the dataflow dependency on annotated function invocations, through tracing the usage of their return values. Further, all accesses to annotated data objects and variables in context are identified. Functions including instructions that uses the return value of @compute calls will be divided at the location that uses the results from @compute calls. Recursive call to @compute is not supported in \sys.

When generate compiled code, For remote invocations, extra code is generated to request the $mmap$ed location of remote objects and get/set variables in context.

\sys adaptively generate different interfaces when materialization, if \sys knows the source and target memory types. If an access is mainly on local memory, a local interface is generated, otherwise, remote accessing interface adapt to access pattern will be generated. For example, for accessing memory over RDMA, \sys will consider the remote memory latency and access pattern using \cite{Mira}. For synchronization, it will be generated using local sync or remote sync implemented by \sys. Certain patterns are required for \sys to generate efficient access interface.

It's possible that a single memory component is launched as multiple physical components to different locations, for example, a compute component is accessing both local and remote memory objects. This could happen before or after the accessing compute component is scheduled, that is, before or after the memory access interface is generated. Both local and remote accessing interface could handle the case: raw memory interfaces accesses remote memory through swapping, and remote interface could access local memory through local cache section described in Mira \cite{Mira}.

% \subsection{Internal Interface}

% The internal interface is fed to \sys's execution system. 
% Thus, we design it in such a way to express explicit component graphs and interactions between components.
% With this internal representation, \sys's execution system does not need to directly analyze high-level language or application features.
% This interface is intended to be used as compiler output or by library and runtime developers.
%
%Users of this interface submit a JSON file and a set of code pieces to \sys.  
% Specifically, the interface is described by a JSON %file that represents a component graph.
%We offer a JSON 
% template that specifies virtual compute and memory components and the relationships between them in an application (\ie, one JSON file represents one component graph). Each virtual compute component points to a code piece Different materialization results are generated and cached at the object time and packed with potential co-running components.

%Optionally, users could give hints about the initial resource amount of an component in the JSON file. %When launching a function, \sys will decide real resource usage for components considering resource hint and execution history of component in the materialization process.
This interface also includes a set of \sys\ low-level APIs as shown in Figure~\ref{fig-api} that a compute component code can use to access other memory components or communicate with other compute components.
%to communication with other components, such as creating a \texttt{channel} to access another component, \texttt{read} and \texttt{write} to an offset of an opened \texttt{channel} to a memory element, \texttt{send} and \texttt{recv} an opened \texttt{channel} between two compute elements, and \texttt{release} an opened \texttt{channel}. %\zhiyuan{first, user need to create the channel between compute element and its targeting memory/compute element. Then it can call a set of APIs based on the channel. 1) a set of communication APIs, for memory elements, read and write. For compute elements, send and recv. 2) a set of. I think its good to list here. }

\section{Additional Design Details}

\subsection{Memory Component Isolation}

On memory side, different memory regions must not share same page and are isolated through the virtual memory system or the paging. 

On accessing memory regions, as discussed in \S\ref{sec:communication}, we support both one-sided remote memory accesses using RDMA and two-sided communication using TCP. 
For RDMA, we assign each physical memory component its own memory region (MR) and own protection domain (PD) for proper isolation. when a memory controller receives a request from the scheduler to start a new physical memory component, it launches one process to allocate this physical memory component's size and registers it with RDMA with a new MR and a new PD.
Afterwards, accesses to the physical memory components are all one-sided operations that do not involve the memory controller (\S\ref{sec:communication}).
For two-sided TCP, we use the memory controller to allocate memory in a global virtual memory space at physical memory component launching time. It then receives/responds memory-accessing messages and copies message data to/from the allocated memory space. As all accesses go through the trusted \sys\ memory controller, physical memory components are properly protected even in one global memory space.

\subsection{Swap System for Compute Components}
Our remote-memory swapping happens entirely in the user space using Linux \ufd\ and is transparent to user applications. Specifically, the \sys\ runtime uses a background thread to monitor page faults caused by the user application threads.
When a fault happens, if there is not enough swap space, the runtime asks the scheduler to create and launch a new physical memory component.
The runtime's background thread swaps out pages whenever it detects memory pressure.
Since the user-space fault handler cannot access the page table and would not know the page access pattern, we use an NRU (not-recently-used) policy by swapping out a page that has not recently been swapped in.

\subsection{Resource adjustment algorithm}
With a goal of minimizing resource waste and maximize the performance at the same time, we model the proactive allocation process as an linear optimization problem. The goal of the algorithm is for each component to select the best initial resource amount $inti$ and increment scaling size $step$.

$$
\begin{aligned}
\min_{step,init} && init + \sum_{h \in History}{step \times k_h \times cost\_factor} && \\
\textrm{s.t.} && h \in History\ |\ k_h \times step + init && > && h \\
&& \frac{\sum_{h \in History}{\max(init - h, 0) \times exec\_time_h}}{\sum_{h \in History}{h}} && < && Thres
\end{aligned}
$$

$cost\_factor$ is a factor that models the scaling cost. We use ortools \cite{ortools} to solve the optimization problem.

\subsection{RDMA-based Communication Control Path}
When establishing an RDMA connection (\ie, RDMA QP) between two nodes, they need to first exchange a set of metadata describing their own identities.
%Today's RDMA systems use TCP to exchange these messages.
% However, two components may not know each other's network addresses due to the dynamic nature of scheduling in a serverless setting.
However, two components cannot easily reach each other or establish a connection to perform this initial metadata exchange due to the dynamic and isolated nature of serverless computing.
Prior solutions either use a costly overlay network layer (which accounts for nearly 40\% of startup time in our exeperiments or require container runtime changes for performance improvement~\cite{thomas2020particle}\footnote{References are in the main submission file.}) or pre-establish all connections~\cite{copik2021rfaas,wei2022booting,wei2021krcore,mohan2019agile} (which does not fit our need to dynamically launch memory components).
Both approaches try to establish connections via {\em direct} channels between two nodes.
%, or require
%Recent RDMA-based serverless systems pre-establish RDMA connections between all the servers in a cluster \cite{copik2021rfaas,wei2022booting,wei2021krcore}, which only works with static memory regions and do not fit our dynamic launched components. 
%Another prior work proposes to maintain a pool of containers with pre-assigned IP addresses~\cite{mohan2019agile}. 
%Unlike \sys, this work still requires TCP connection to be setup when launching a function.
%Moreover, reusing IP addresses could break the isolation between functions.
%An alternative dynamic approach is to use an overlay network to enable direct %communication between containers. 
%However, our experiments show that the initialization of the overlay network takes nearly 40\% of the startup time of a container.
%\zhiyuan{removed overlay here}
%A recent work~\cite{thomas2020particle} enables faster overlay network creation, but requires changes to the container runtime code. % and is not available publicly.

%\zhiyuan{reouting message in a pre-setup network between executors and schedulers instead of direct communcation between elements.}
Our idea is to leverage {\em non-direct but already established} connections between executors and schedulers to exchange the initial metadata message.
Our observation is that a component has always established a connection with its rack-level scheduler by the time when it starts up.
Moreover, the scheduler is the one that decides and thus knows the physical locations of the two components that will be communicating. % (\ie, communicating relationship in the component graph).
We let the scheduler send one side's (\eg, $B$'s) physical location (in the form of executor ID) to the other side's (\eg, $A$'s) {\em network virtualization module} when initiating the components.
%This network module sits inside the executor and only manages connection (RDMA QP) establishment.
%After receiving the destination component's executor ID, 
Afterwards, $A$'s network virtualization module sets up and sends the necessary RDMA connection metadata of $A$ together with the destination component's executor ID ($B$) to the scheduler.
%When a component wants to establish QP with another component, it sends the metadata message to the scheduler with the other component's ID.
The scheduler routes the message containing $A$'s metadata to the target executor, which then gets sent to the destination component ($B$) by its network module.
\if 0
When a component wants to establish QP with another component, it sends the metadata message to the scheduler with the other component's executor ID.
The scheduler forwards the message to this executor ID, which then gets sent to this other component by its network module.
\fi
After both sides acquire the other's metadata message, they establish an RDMA QP. The entire QP establishment takes only 34\ms.

To further optimize the startup performance, the QP establishment process starts as soon as the execution environment is ready, while user code is loaded in parallel. Doing so hides the QP establishment overhead behind the performance-critical path.
%\zhiyuan{Async setup before the first communication happens. }
%Another idea we have is to {\em asynchronously} prepare and establish RDMA QP so that user logic can start before QP is established. We will detailed this design in \S\ref{sec:coldstart}.

Furthermore, we reuse an RDMA QP when a component tries to establish the communication with a physical memory component located on a server that already has another physical memory component communicating with the component. Since the new and the existing physical memory components are both accessed by the same component, there is no need to isolate them and one QP is enough for both components. 

\subsection{RDMA-based Communication Data Path}
\sys's RDMA communication data path bypasses kernel.
When a compute component accesses another component (compiled as an internal communication API),
%calls a low-level communication API through \texttt{channel},
%(\eg, \texttt{read}, \texttt{write}, \texttt{send}, \texttt{recv})
the \sys\ runtime looks up which QP it corresponds to. 
It then issues a one-sided RDMA operation for \texttt{read}/\texttt{write} for memory components or a two-sided RDMA operation for \texttt{send}/\texttt{recv} for compute components. 
As one virtual component can be mapped to multiple physical components,
if the internal API call touches a memory region that is on two physical memory components, the \sys\ runtime dispatches two one-sided RDMA operations to these components and merge the responses.
%and re-assemble the two results before returning the call.
%for each targeted physical memory component and sends out these RDMA operations via the corresponding QPs in parallel.
%After receiving data from multiple RDMA operations, it 
% To optimize communication performance and avoid serialization cost, 
We exploit zero-copy techniques when implementing the data path to improve performance.
%the low-level APIs and the high-level library support.
%\yiying{Zhiyuan, do we do zero copy for the raw API interface? for the numpy/dataframe interface? How do we do that?} \zhiyuan{yes we do. For advanced usages we also exposes raw rdma buffer interface.}

\section{Additional Results}

% {
% \begin{figure}[t]
% \begin{center}
% \centerline{\includegraphics[width=\columnwidth]{Figures/Figure-merge-appendix.pdf}}
% % \centerline{\includegraphics[width=0.97\columnwidth]{Figures/Figure-policy-line.pdf}}
% \vspace{-0.1in}
% \end{center}
% \vspace{-0.2in}
% %caption
% \captionsetup{width=0.225\textwidth}
% \begin{minipage}{0.49\columnwidth}
% \begin{center}
% \mycaption{fig-micro-swap}{\sys Performance with Swap Component.}{}
% \end{center}
% \end{minipage}
% \begin{minipage}{0.49\columnwidth}
% \begin{center}
% \mycaption{fig-appendix-policy}{Resource Distribution in Adjustment Experiment.}{}
% \end{center}
% \end{minipage}

% \vspace{-0.1in}
% \end{figure}

{
\begin{figure}[t]
\begin{center}
\centerline{\includegraphics[width=\columnwidth]{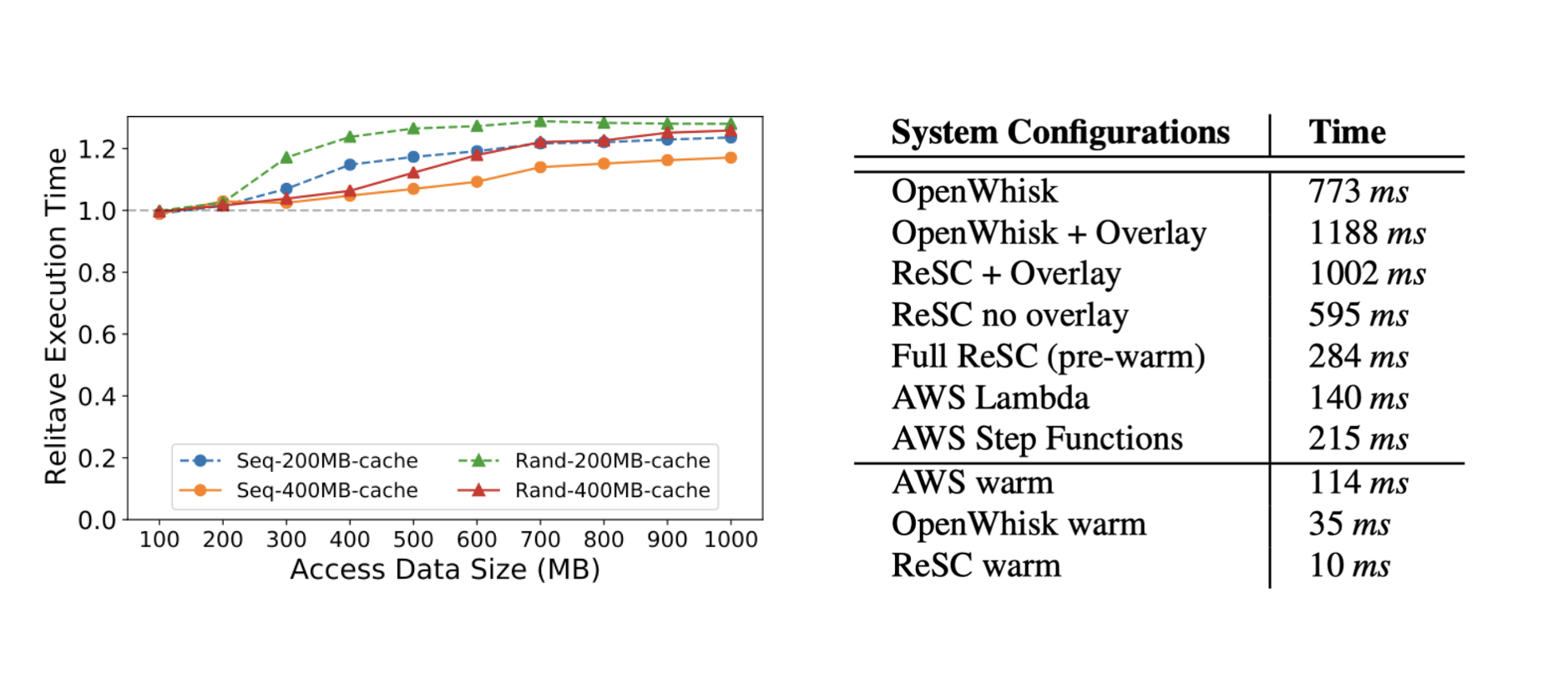}}
\vspace{-0.1in}
\end{center}
\vspace{-0.2in}
%caption
\captionsetup{width=0.225\textwidth}
\begin{minipage}{0.49\columnwidth}
\begin{center}
\mycaption{fig-micro-swap}{\sys Performance with Swap Component.}{}
\end{center}
\end{minipage}
\begin{minipage}{0.49\columnwidth}
\begin{center}
\mycaption{fig-appendix-policy}{Cold and warm start up time}{}
\end{center}
\end{minipage}

\vspace{-0.1in}
\end{figure}
}
{
\begin{figure}[t]
\begin{minipage}{0.49\columnwidth}
\begin{center}
\centerline{\includegraphics[width=\columnwidth]{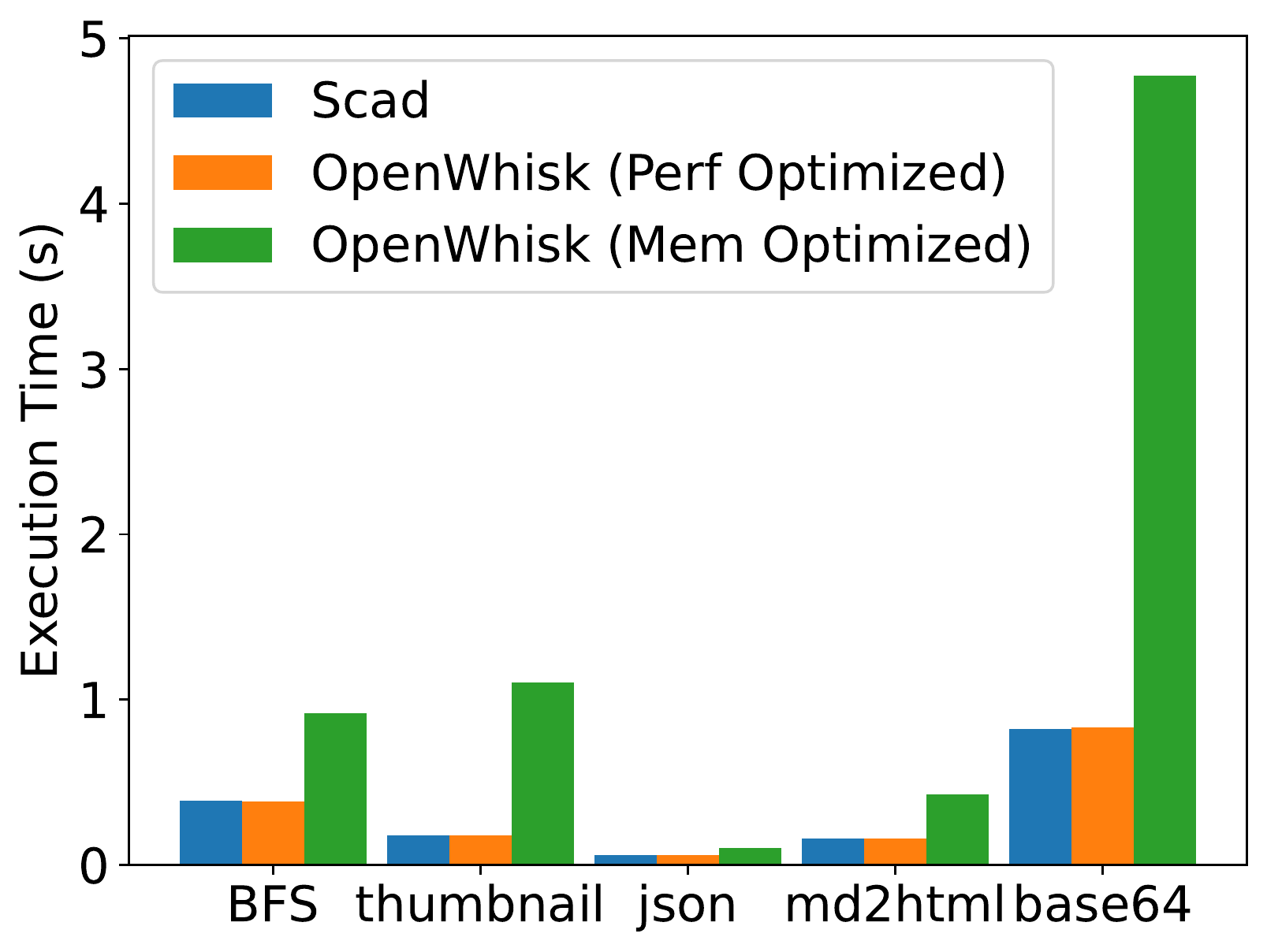}}
\mycaption{fig-micro-time}{\sys Execution Time on Small Applications.}{}
\end{center}
\end{minipage}
\begin{minipage}{0.49\columnwidth}
\begin{center}
\centerline{\includegraphics[width=\columnwidth]{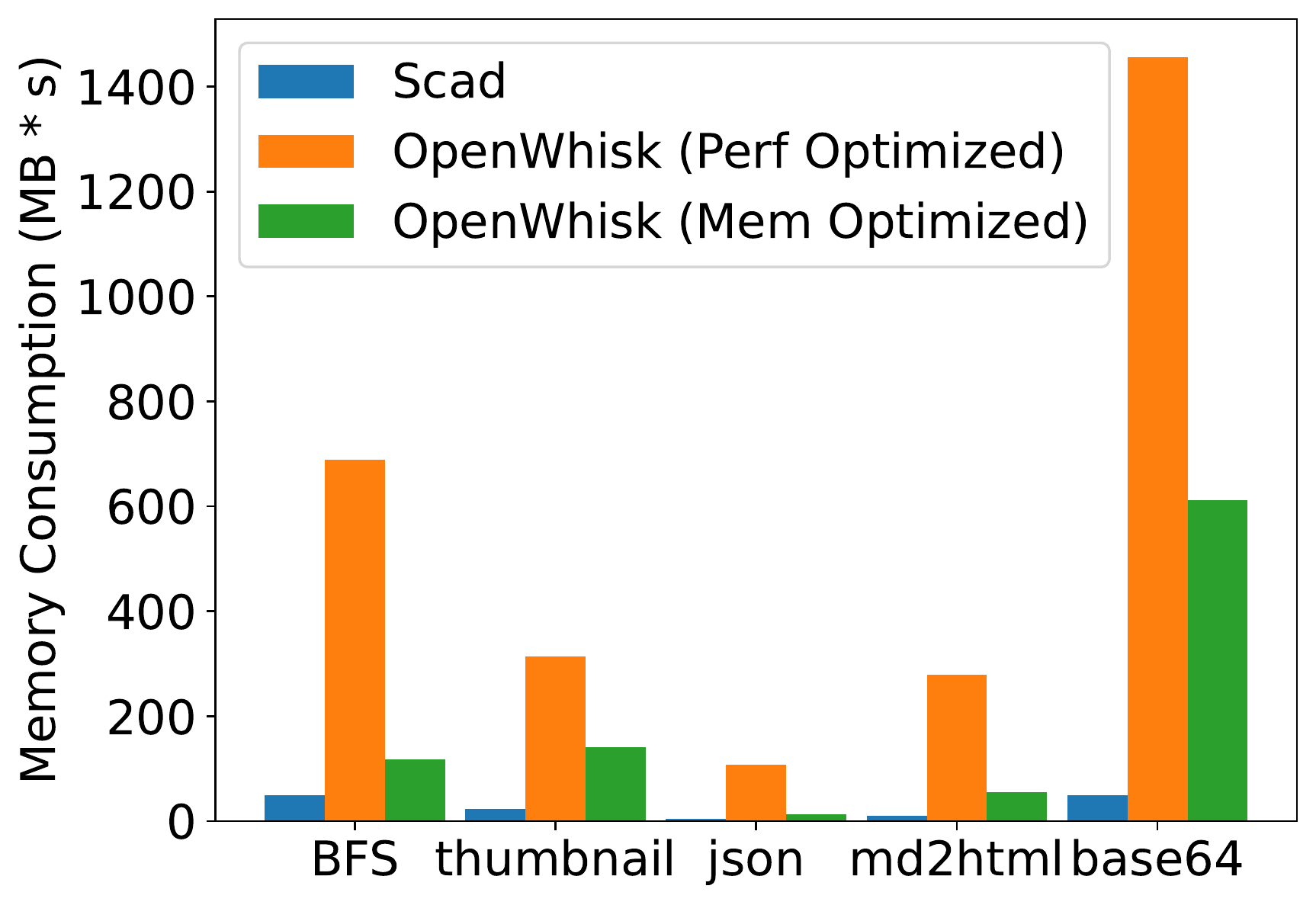}}
\mycaption{fig-micro-resource}{\sys Resource Consumption on Small Applications.}{}
\end{center}
\end{minipage}
\end{figure}

We provide four sets of additional evaluation results.

\boldunderpara{Auto-scaled compute components with memory swapping}
%\sys\ automatically and transparently scales a compute component up with memory swapping when the component is low on memory. 
To measure our swap system performance, we use a simple microbenchmark of sequentially or randomly reading an array of different sizes.
Figure~\ref{fig-micro-swap} plots the total run time of the microbenchmark with increasing array size. 
We test two local cache sizes: 200\MB\ and 400\MB, and compare the swap performance against the case where the local memory is larger than the whole array size (\ie, no swapping and the ideal performance but impossible with the compute pool configuration).
Overall, swapping only adds 1\% to 26\% performance overhead. The overhead is higher when the array size is bigger and when the local memory size is smaller.

\boldunderpara{Scheduler Scalability}
%When applications are split causing more components to be created, more load will be applied to \sys's scheduler. 
We evaluate the maximum throughput of \sys' rack-level scheduler and find that it can handle around 20K scheduling requests per second.
%a single rack scheduler in \sys. To measure this we applied load by sending a large number of scheduling messages from all machines across our system. A single rack scheduler is able to achieve a maximum throughput of 20k scheduling messages per second. 
This scheduling rate is similar to OpenWhisk's scheduler. However, thanks to our locality based design with a scheduler-per-rack, the number of total messages that the system can handle scales linearly with the number of racks.
We also find the request rate to the top-level scheduler to be light (around 50K requests per second) and is never the bottleneck in the system. 
%\yiying{Zac, do we have any numbers for the request rate?} \zacb{I don't have numbers recorded but if I recall we were able to drive ~50k rps from the clients. I didn't go any further than that because I knew the scheduler wasn't keeping up. I would have to go back and probably run the code again to see exactly what the number is.}

\boldunderpara{Per-input component adjustment.} We present more details of the workloads used for evaluating per-input component adjustment (\S6.2.3) Figure \ref{fig-appendix-policy} shows the four selected applications' memory utilization distribution across invocations.  \textbf{Large} and \textbf{Small} features a large or small average memory, which is far different from our default setup. \textbf{Varying} features a relative high variance, which can be representative for applications with varying resource requirements. \textbf{Stable} in contrast, features almost the same resource utilization across all invocations.
For comparison, we also plot a straight line of 256\MB, which is the default initial memory allocation size in \sys.

\boldunderpara{Small Application Performance.} 
Figures~\ref{fig-micro-time} and~\ref{fig-micro-resource} 
showcase the execution time and resource consumption of five small serverless functions from SeBS~\cite{copik2021sebs} and FaaSProfiler~\cite{Shahrad_Micro19} benchmarks.
These single-function applications have an execution time of less than one second and memory usage smaller than 128\MB.
Although these applications do not benefit from \sys's resource-centric scaling, \sys still outperforms OpenWhisk in terms of resource consumption while delivering similar performance. 
This is attributed to \sys's ability to flexibly allocate resources.

\boldunderpara{Solver Performance.} 
We implemented the solver using the highly efficient Python Mixed-Integer Linear Program (MIP) package.
Finding the optimal solution for 10000 disaggregation candidates each with 32 components takes $10ms$-$15ms$.

% New figure about cluster level utilization
\boldunderpara{Performance/cost on a fixed cluster.} 
\sys's resource saving reflects at the cluster level. We evaluate \sys and Openwhisk with the same total cluster resources as shown in Figure~\ref{fig:eval:cluster-mem}.
First, \sys could allocate more resources compared to resource-tuned serverless, because of resource-level scaling reduces resource stranding and utilizes free resources across server; Second, most of allocated resources are utilized in \sys, due to the runtime resource scaling adapts to resource size. As a result, given same amount of resources, \sys could utilize more resources in real computing, results in a 33\% to 90\%  performance gain compared to Openwhisk.

\begin{figure}[t]
\begin{minipage}{0.48\columnwidth}
\begin{center}
%\centerline{\includegraphics[width=\columnwidth]{Figures/fig_tpcds_stage.pdf}}
\centerline{\includegraphics[width=\columnwidth]{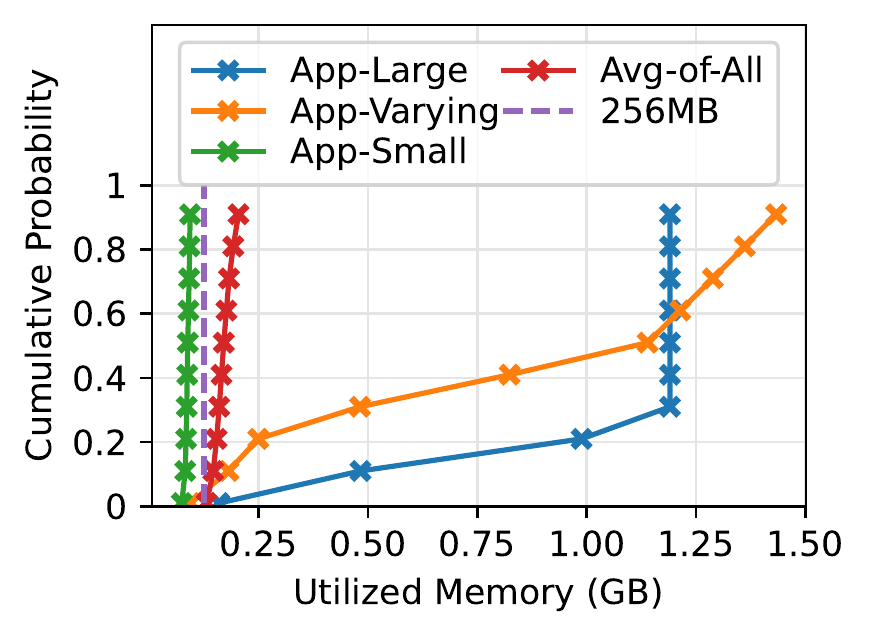}}
\vspace{-0.1in}
%\mycaption{fig-tpcds-breakdown}{Breakdown of Different Versions of TPCDS.}
\mycaption{fig-dataset-dist}{Memory Usage Distribution of Azure Dataset.}
{
}
\end{center}
\end{minipage}
\hspace{0.015\columnwidth}
\begin{minipage}{0.48\columnwidth}
\begin{center}
\centerline{\includegraphics[width=\columnwidth]{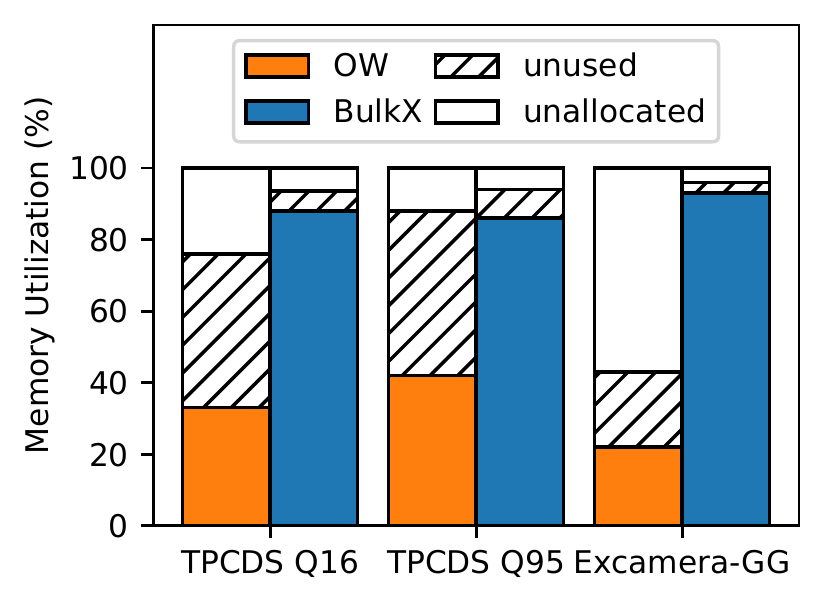}}
\mycaption{fig:eval:cluster-mem}{Cluster Level Memory Utilization}
{
}
\end{center}
\end{minipage}
\end{figure}

\end{document}